\documentclass[aps,prd,10pt,preprint,superscriptaddress,onecolumn,nofootinbib]{revtex4-1}
\usepackage{graphicx, epsfig} 
\usepackage{amsmath,amssymb,amsfonts,dsfont,mathrsfs,amsthm,mathtools}
\usepackage{bm} 
\usepackage{color}
\usepackage[usenames]{xcolor}
\usepackage{hyperref}
\usepackage{siunitx}
\hypersetup{colorlinks=true,urlcolor=blue,linkcolor=magenta,citecolor=blue,filecolor=blue}
\usepackage[normalem]{ulem}
\usepackage{array}
\usepackage{hyperref}
\usepackage{booktabs}
\usepackage{blindtext}

\newcommand{\diff}[1]{\text{d}#1}

\newcommand{\Lag}{\mathscr{L}}

\begin{document}

\title{Conformal Renormalization of topological black holes in AdS$_6$}

\author{Giorgos Anastasiou}
\email{ganastasiou@unap.cl}
\affiliation{Instituto de Ciencias Exactas y Naturales, Universidad Arturo Prat, Playa Brava 3256, 1111346, Iquique, Chile}
\affiliation{Facultad de Ciencias, Universidad Arturo Prat, Avenida Arturo Prat Chac\'on 2120, 1110939, Iquique, Chile}

\author{Ignacio J. Araya}
\email{ignaraya@unap.cl}
\affiliation{Instituto de Ciencias Exactas y Naturales, Universidad Arturo Prat, Playa Brava 3256, 1111346, Iquique, Chile}
\affiliation{Facultad de Ciencias, Universidad Arturo Prat, Avenida Arturo Prat Chac\'on 2120, 1110939, Iquique, Chile}
\affiliation{School of Mathematics and Hamilton Mathematics Institute, Trinity College, Dublin 2, Ireland}

\author{Crist\'obal Corral}
\email{crcorral@unap.cl}
\affiliation{Instituto de Ciencias Exactas y Naturales, Universidad Arturo Prat, Playa Brava 3256, 1111346, Iquique, Chile}
\affiliation{Facultad de Ciencias, Universidad Arturo Prat, Avenida Arturo Prat Chac\'on 2120, 1110939, Iquique, Chile}

\author{Rodrigo Olea}
\email{rodrigo.olea@unab.cl}
\affiliation{Universidad Andres Bello, Departamento de Ciencias F\'isicas, \\  Facultad de Ciencias Exactas, Sazi\'e 2212, Piso 7, Santiago, Chile}

\begin{abstract}
 We present a streamlined proof that any Einstein-AdS space is a solution of the Lu, Pang and Pope conformal gravity theory in six dimensions. The reduction of conformal gravity into Einstein theory manifestly shows that the action of the latter can be written as the Einstein-Hilbert term plus the Euler topological density and an additional contribution that depends on the Laplacian of the bulk Weyl tensor squared. The prescription for obtaining this form of the action by embedding the Einstein theory into a Weyl-invariant purely metric theory, was dubbed \emph{Conformal Renormalization} and its resulting action was shown to be equivalent to the one obtained by holographic renormalization. As a non-trivial application of the method, we compute the Noether-Wald charges and thermodynamic quantities for topological black hole solutions with generic transverse section in Einstein-AdS$_6$ theory.
\end{abstract}

\maketitle

\section{Introduction\label{sec:intro}}









Conformal Gravity (CG)~\cite{Weyl:1918ib,Weyl:1919fi,Bach:1921} is a higher-derivative theory in four dimensions that has been extensively studied as an alternative to General Relativity. Indeed, it has a better ultraviolet behavior~\cite{Stelle:1976gc,Capper:1975ig,Fradkin:1981iu,Julve:1978xn} and can accommodate astrophysical effects such as the flat galaxy rotation curves without the need of dark matter~\cite{Mannheim:1988dj,Mannheim:2005bfa,Mannheim:2010ti,Mannheim:2011ds}. The origins of CG can be found in twistor string theory~\cite{Berkovits:2004jj}, what makes it a useful tool for the construction of supergravity theories~\cite{Kaku:1978ea,Kaku:1978nz,Kaku:1977pa,Fradkin:1985am,deWit:1980lyi,Bergshoeff:1980is,Liu:1998bu,Ferrara:2018wqd,Andrianopoli:2014aqa,DAuria:2021dth,Ferrara:2020zef} and non-commutative geometries~\cite{Chamseddine:1996zu,Manolakos:2019fle,Manolakos:2021rcl}. However, as it violates the Lovelock theorem, it introduces Ostrogradsky instabilities~\cite{Ostrogradsky:1850fid} which generically lead to ghost-like degrees of freedom~\cite{Riegert:1984hf}. This fact does not preclude the possibility of the theory having no ghosts when linearized about certain backgrounds, as mentioned in Ref.~\cite{Mannheim:2021oat}.

Interestingly, in the context of the anti-de Sitter/Conformal Field Theory (AdS/CFT) correspondence, a holographic mechanism to obtain  Einstein gravity from CG was proposed by Maldacena in Ref.~\cite{Maldacena:2011mk}. There, the former theory can be obtained from the latter upon imposing Neumann boundary conditions in the  metric near the conformal boundary of AdS.  A more explicit analysis was carried out in Ref.~\cite{Grumiller:2013mxa} where the non-Einstein holographic source and its corresponding response were identified for CG in four dimensions.  As a matter of fact, in the Feffermann-Graham (FG) frame~\cite{AST_1985__S131__95_0}, the new higher-derivative mode acts as the source of the partially massless current associated to the massive graviton present in the CG spectrum.  Therefore, the Neumann boundary condition discussed in Ref.~\cite{Maldacena:2011mk} is equivalent to turning off the additional holographic source. Then, the expansion of the metric reduces to that of the Einstein sector of the theory. In a similar way, one recovers Einstein-de Sitter spacetimes from CG imposing a Neumann condition at the future timelike boundary, whereas a set of boundary conditions has been proposed that permits the reduction to asymptotically flat Einstein spacetimes~\cite{Hell:2023rbf}.

As regards the bulk action, it was shown in Ref.~\cite{Anastasiou:2016jix} that when Einstein spacetimes are considered, the CG action reduces to that of Einstein-AdS in its MacDowell-Mansouri form~\cite{MacDowell:1977jt}, i.e., written in terms of the square of the AdS curvature. This is obtained from the Weyl-squared invariant when non-Einstein degrees of freedom are turned off, thereby breaking conformal symmetry. Surprisingly enough, this bulk functional is already renormalized and equivalent to what is obtained through the standard holographic renormalization  procedure~\cite{Miskovic:2009bm}.

The relation between Einstein-AdS gravity and CG is therefore twofold. Firstly, at the level of the equations of motion (EOM), all solutions to Einstein-AdS gravity are also solutions to CG. Secondly, its action becomes renormalized Einstein-AdS gravity when evaluated on Einstein spaces. This scheme corresponds to a consistent embedding of Einstein-AdS gravity into CG. In this way, the Einstein sector inherits the renormalization properties of the conformally invariant theory~\cite{Grumiller:2013mxa}, making it free of infrared (IR) divergences coming from the conformal boundary of asymptotically AdS (AAdS) spaces.

This idea was generalized to six dimensions in Ref.~\cite{Anastasiou:2020mik}. Indeed, there exists a particular combination of the three conformal invariants in 6D which defines a CG theory that admits any Einstein space as a solution~\cite{Lu:2013hx}. In Ref.~\cite{Anastasiou:2020mik}, the FG expansion of the metric was considered and the holographic conditions necessary to recover Einstein-AdS gravity were identified.  These  constraints  on the asymptotic form of the metric naturally extend the Neumann condition in Ref.~\cite{Maldacena:2011mk} and provide a holographic mechanism to obtain the renormalized Einstein-AdS gravity in six dimensions. In point of fact, the renormalization prescription  obtained from CG in this way supersedes the one defined by the addition of bulk topological terms to the Einstein-Hilbert action with negative cosmological constant \cite{Olea:2005gb}. In particular, the resulting boundary terms act as surface counterterms which match the one given by the standard holographic renormalization procedure~\cite{deHaro:2000vlm,Henningson:1998gx}.  The cancellation of divergences in the bulk AdS gravity action by embedding it into a conformally invariant  theory is referred to as \emph{Conformal Renormalization}. 

In this work, we apply the aforementioned prescription to the case of Einstein-AdS gravity considering solutions whose radial hypersurfaces are not conformally flat. In particular, we focus on topological AdS black holes (BHs) with $\mathbb{S}^2\times\mathbb{S}^2$, $\mathbb{H}^2\times\mathbb{H}^2$, $\mathbb{CP}^2$ and $\mathbb{CH}^2$ transverse sections. We verify the finiteness of the Euclidean on-shell action and of the conserved asymptotic charges, thus testing the consistency of the method.\footnote{The Conformal Renormalization scheme has also been applied to scalar-tensor theories in Ref.~\cite{Anastasiou:2022wjq}.}

The paper is organized as follows. In section~\ref{TopologicalRenormalization} we introduce the topological renormalization prescription for Einstein-AdS gravity in six dimensions. There we provide a detailed analysis of the asymptotic properties of the geometric objects involved in the corresponding action. In section~\ref{ConformalRenormalization} we determine the renormalized Einstein-AdS action in six dimensions using Conformal Renormalization. In section~\ref{sec:Schw} we provide an explicit application of this prescription to topological AdS BHs with different transverse sections and its comparison to holographic renormalization. Finally, in section~\ref{NW}, we calculate the corresponding Noether-Wald charges and thermodynamic properties of the aforementioned solutions.

\section{Topological Renormalization of Einstein-AdS gravity in six dimensions}\label{TopologicalRenormalization}

In this section, we consider the renormalization of the six-dimensional Einstein-Hilbert action by augmenting it with a boundary term related to the topological invariant of the Euler class. In order to fix this prescription, we first note that in $D=2m$ dimensions, the Euler theorem states that 
\begin{align}\label{Eulertheorem}
   \int_{\mathcal{M}_{2m}}\diff{^{2m}x}\sqrt{|g|}\, \mathcal{E}_{2m} = \left( 4\pi\right)^m\,m!\,\chi\left(\mathcal{M}_{2m}\right) + \int_{\partial\mathcal{M}_{2m}}\diff{^{2m-1}x}\sqrt{|h|}\,B_{2m-1}\,,
\end{align}
where $\mathcal{E}_{2m}$ is defined as
\begin{align}
 \mathcal{E}_{2m} = \frac{1}{2^m}\delta^{\mu_1\ldots\mu_{2m}}_{\nu_1\ldots\nu_{2m}}R^{\nu_1\nu_2}_{\mu_1\mu_2} \ldots R^{\nu_{2m-1}\nu_{2m}}_{\mu_{2m-1}\mu_{2m}}\,,
\end{align}
and $\chi\left(\mathcal{M}_{2m}\right)$ is the Euler characteristic of the manifold $\mathcal{M}_{2m}$. Here, $B_{2m-1}$ is a boundary term present for non-compact manifolds with boundaries, dubbed the $m-$th Chern form. 
For our purposes we shall take a radial foliation of the spacetime, given by
\begin{equation}\label{GaussNormal}
\diff{s^2}=N^2(z)\diff{z^2}+h_{ij}(z,x)\diff{x^i} \diff{x^j}\,,
\end{equation}
with $z=0$ indicating the location of the conformal boundary. In this coordinate frame, the Chern form is written as
\begin{align}\notag
    B_{2m-1} &= 2m\int_0^1\diff{s}\,\delta^{i_1\ldots i_{2m-1}}_{j_1\ldots j_{2m-1}}K^{j_1}_{i_1}\left(\frac{1}{2}\mathcal{R}^{j_2j_3}_{ i_2 i_3} - s^2\,K^{j_2}_{i_2}K^{j_3}_{i_3} \right)\times \\
    &\quad \ldots\times\left(\frac{1}{2}\mathcal{R}^{j_{2m-2}j_{2m-1}}_{ i_{2m-2}i_{2m-1}} - s^2\,K^{j_{2m-2}}_{i_{2m-1}}K^{j_{2m-1}}_{i_{2m-1}} \right)\,,
\end{align}
where $\mathcal{R}^{ij}_{kl}$ and $K_{ij}=-\frac{1}{2N} \partial_{z} h_{ij}$ are the intrinsic and extrinsic curvatures, respectively, related to the Riemannian curvature through the Gauss-Codazzi equation
\begin{align}
    \mathcal{R}^{ij}_{kl} =  R^{ij}_{kl} + K^i_{k} K^{j}_{l} - K^i_{l} K^{j}_{k}  \,.
\end{align}

Topological renormalization dictates the addition of the five-dimensional Chern form to the Einstein-AdS action, such that we get
\begin{align}
 I_{\rm EH} &= \kappa\int_{\mathcal{M}_{6}}\diff{^6x}\sqrt{|g|}\left(R+\frac{20}{\ell^2} \right)+\kappa \eta \int_{\mathcal{\partial M}_{6}}\diff{^5x}\sqrt{|h|}B_{5} \,.
\end{align}
Here, $\kappa=(16\pi G)^{-1}$ is the gravitational constant, $\ell$ is the AdS radius related to the cosmological constant through $\Lambda=-\tfrac{10}{\ell^2}$, and $\eta$ is the coupling constant of the five-dimensional Chern form.
Finally, by virtue of the Euler theorem of Eq.~\eqref{Eulertheorem}, the renormalized action can be equivalently written as
\begin{align}\label{EH}
I_{\rm EH} &= \kappa\int_{\mathcal{M}_{6}}\diff{^6x}\sqrt{|g|}\left(R+\frac{20}{\ell^2} + \eta \mathcal{E}_6 \right)+\tilde{\chi}_6 \,,
\end{align}
where
\begin{align}\label{chitilde}
\tilde{\chi}_6=-6\kappa \eta (4\pi)^{3}\chi(\mathcal{M}_{6}) \,,
\end{align}
is a topological number which is proportional to the Euler characteristic of the bulk manifold $\mathcal{M}_{6}$. 
For convenience, we will express this topologically-renormalized Einstein-AdS action in terms of the Euler density instead of the Chern form. Note that the $\tilde{\chi}_6$ does not play a role in determining the dynamics of the theory, but it does give a constant contribution to the entropy of BH solutions.\footnote{Also, in a holographic context, this term gives the finite topological contribution to the entanglement entropy of spherical regions, as discussed in Refs.~\cite{Anastasiou:2018mfk,Anastasiou:2019ldc}.}

The field equations are obtained by demanding stationary variations of the action~\eqref{EH} with respect to the metric, giving the Einstein field equations in six dimensions; they are,
\begin{align}\label{EinsteinEq}
 R_{\mu\nu} - \frac{1}{2}g_{\mu\nu}R - \frac{10}{\ell^2} g_{\mu\nu} = 0\,.
\end{align}
Furthermore, the Weyl tensor evaluated on Einstein spaces can be written as
\begin{align}\label{WeylE}
  W^{\mu\nu}_{{\rm (E)}\lambda\rho}= R^{\mu\nu}_{\lambda\rho} + \frac{1}{\ell^2}\delta^{\mu\nu}_{\lambda\rho},
\end{align}
which equals the torsion-free piece of the curvature associated to the AdS group. 

The action~\eqref{EH} is finite for Asymptotically Conformally Flat (ACF)\footnote{These are spacetimes whose bulk Weyl tensor falls-off at the normalizable order.} Einstein spaces if and only if the value of the Euler density coupling satisfies
\begin{align}\label{eta}
 \eta = -\frac{\ell^4}{72}\,.
\end{align}
For this particular choice, the topologically renormalized Einstein-Hilbert action~\eqref{EH} can be written as
\begin{align}\label{EH2}
 I_{\rm EH} &= \kappa\ell^4\int_{\mathcal{M}_6}\diff{^6x}\sqrt{|g|}\,P_6\left(W_{\rm (E)}\right)+\tilde{\chi}_6\,,
\end{align}
where $P_6\left(W_{\rm (E)}\right)$ denotes the polynomial of contractions of the Weyl tensor evaluated on Einstein spaces given by~\cite{Anastasiou:2018mfk}
\begin{align}
    P_6 (W_{\rm (E)} ) = \frac{1}{2\left(4!\right) \ell^2} \delta_{\mu_{1} \ldots \mu_{4}}^{\nu_{1}\ldots \nu_{4}} W_{\left(E\right)\nu_{1} \nu_{2}}^{\mu_{1} \mu_{2}} W_{\left(E\right)\nu_{3} \nu_{4}}^{\mu_{3} \mu_{4}} -\frac{1}{\left(4!\right)^2} \delta_{\mu_{1} \ldots \mu_{6}}^{\nu_{1} \ldots \nu_{6}} W_{\left(E\right)\nu_{1} \nu_{2}}^{\mu_{1} \mu_{2}} W_{\left(E\right)\nu_{3} \nu_{4}}^{\mu_{3} \mu_{4}} W_{\left(E\right)\nu_{5} \nu_{6}}^{\mu_{5} \mu_{6}}\,.
    \label{P6WE}
\end{align}
In what follows, we analyze the asymptotic behavior of $P_6\left(W_{\rm (E)}\right)$ and determine the class of manifolds for which the action~\eqref{EH2} is free of IR divergences.

\subsection*{Asymptotic analysis of $P_{6} \left(W_{\left(E\right)}\right)$}


Due to the radial foliation, there are three independent components of the Weyl tensor whose fall-off needs to be analyzed, i.e., $W_{ms}^{ij}$, $W_{jm}^{iz}$ and $W_{zj}^{zi}$;  the latter can be identified as its electric part $E_{j}^{i}$. At this point, it is more convenient to work out the expressions of Eq.~\eqref{GaussNormal} considering the FG expansion of the metric for Einstein spaces, that is,
\begin{equation}
\diff{s}^2=\frac{\ell^2}{z^2} \left(\diff{z}^2+\tilde{g}_{ij}\left(z,x\right)\diff{x}^{i}\diff{x}^{j}\right)\,,
\end{equation}
with
\begin{equation}
\tilde{g}_{ij}\left(  z,x\right)  =g_{\left(  0\right)  ij}\left(  x\right)    +\frac{z^{2}}{\ell^{2}%
}g_{\left(  2\right)  ij}\left(  x\right)  +\frac{z^{4}}{\ell^{4}}g_{\left(  4\right)
ij}\left(  x\right)  +\frac{z^{5}}{\ell^{5}}g_{\left(  5\right)  ij}\left(
x\right)  +\mathcal{O}\left(  z^{6}\right)  \,.
\label{6DFGexpansion}
\end{equation}
Taking into account this expansion, the fall-off of the components of the Weyl tensor reads
\begin{subequations}
\begin{align}
E_{\left(\text{E}\right)j}^{i} &= z^{4} \mathcal{B}_{j}^{i} \left(g_{(0)}\right)+\mathcal{O} \left(z^5\right) \,, \\ W_{\left(\text{E}\right)ms}^{ij} &=z^{2} \mathcal{W}_{ms}^{ij} \left(g_{(0)}\right) +\mathcal{O} \left(z^4\right) \,, \\W_{\left(\text{E}\right)jm}^{iz} &= z^{3} \mathcal{C}_{jm}^{i} \left(g_{(0)}\right)+\mathcal{O} \left(z^5\right) \,,
\label{weylfalloff}
\end{align}
\end{subequations}
where we have denoted
\begin{align}
B_{\mu}^{\nu}&=\nabla^{\sigma}C^{\mu}_{\nu \sigma}+S^{\sigma}_{\lambda}W^{\mu \lambda}_{\nu \sigma} \,, &
C_{\mu\nu\lambda}&= 2\nabla_{[\lambda} S_{\nu]\mu} \,, &
S^{\mu}_{\nu} &=\frac{1}{D-2}\left[R^{\mu}_{\nu}-\frac{1}{2\left(D-1\right)}R \delta^{\mu}_{\nu}\right] \,,
\label{BachCottonSchouten}
\end{align}
as the Bach, Cotton, and Schouten tensors, respectively. Then, it is straightforward to see that the Pfaffian of the Weyl tensor evaluated on Einstein spaces is free of IR divergences. Indeed, the latter can be defined as
\begin{equation}
\text{Pff}\left
(W_{\left(\text{E}\right)}\right) = \delta_{\mu_{1} \ldots \mu_{6}}^{\nu_{1} \ldots \nu_{6}} W_{\left(\text{E}\right) \nu_{1} \nu_{2}}^{\mu_{1} \mu_{2}} W_{\left(\text{E}\right)\nu_{3} \nu_{4}}^{\mu_{3} \mu_{4}} W_{\left(\text{E}\right)\nu_{5} \nu_{6}}^{\mu_{5} \mu_{6}} = 32 \left(2  I_{\rm CG}^{(1)} + I_{\rm CG}^{(2)} \right)\,,
\label{Pfaffian}
\end{equation}
where $ I_{\rm CG}^{(1)}$ and $ I_{\rm CG}^{(2)}$ are two, out of the three, linearly independent conformal invariants in 6D, as it can be seen in Eq.~\eqref{conformalinvariants} below. In order to perform the asymptotic analysis, it is convenient to write the conformal invariants in terms of contractions of the radial components of the Weyl tensor as
\begin{align}
I_{\rm CG}^{(1)}&= 6 E_{\left(\text{E}\right)m}^{i}W_{\left(\text{E}\right)jz}^{km}\left(  g^{js}g_{kl}W_{\left(\text{E}\right)si}^{lz}\right)
+3 E_{\left(\text{E}\right)m}^{i} E_{\left(\text{E}\right)k}^{j}W_{\left(\text{E}\right)ji}^{mk}+6W_{\left(\text{E}\right)mk}^{ji}W_{\left(\text{E}\right)jz}^{ks}\left(
g_{it}g^{mq}W_{\left(\text{E}\right)qs}^{tz}\right)  \nonumber \\
&+W_{\left(\text{E}\right) ijmk}W_{\left(\text{E}\right)}
^{\ \ islk}W_{\left(\text{E}\right) s~\ \ \  l}^{~\ \ \ jm} \,,\\
\notag
I_{\rm CG}^{(2)}&=8 E_{\left(\text{E}\right)j}^{i} E_{\left(\text{E}\right) m}^{j}E_{\left(\text{E}\right) i}^{m}+12 E_{\left(\text{E}\right)j}^{i}
W_{\left(\text{E}\right) ms}^{jz}W_{\left(\text{E}\right) iz}^{ms}+6W_{\left(\text{E}\right) sk}^{jm}W_{\left(\text{E}\right) iz}^{sk}W_{\left(\text{E}\right)jm}^{iz} \\ &+W_{\left(\text{E}\right)ms}^{ij}W_{\left(\text{E}\right)kl}
^{ms}W_{\left(\text{E}\right)ij}^{kl}\,.
\end{align}
When particularized for Einstein spaces, the fall-off of these objects is of order $\mathcal{O} \left(z^6\right)$. This means that the Pfaffian behaves to the leading order as
\begin{equation}
\text{Pff}\left (W_{\left(\text{E}\right)}\right) \sim \mathcal{O} \left(z^6\right) \,,
\end{equation}
and for the corresponding term in Eq.~\eqref{EH2}, we get
\begin{gather}
\int_{\mathcal{M}_6}\diff{^6x}\sqrt{|g|} \,\text{Pff}\left (W_{\left(\text{E}\right)}\right) \sim \int \diff{z} \int_{\partial\mathcal{M}_6}\diff{^5x} \frac{\sqrt{|g_{\left(0\right)}|}\ell^6}{z^6} z^6 = \text{finite}+ \mathcal{O} \left(z\right) \,.
\end{gather}
Then, the Pfaffian of the Weyl tensor evaluated on Einstein spaces does not contribute with any divergent term in the topologically renormalized Einstein-AdS action. However, as noticed in Ref.~\cite{Anastasiou:2020zwc}, the action of Eq.~\eqref{EH2} is divergent for non-ACF manifolds. As a consequence, the origin of this divergence can be traced back to the quadratic term in the Weyl tensor of Eq.~\eqref{WeylE}. Indeed, this term can be decomposed as
\begin{equation}
\left\vert W_{\left(\text{E}\right)} \right\vert ^{2}=W_{\left(\text{E}\right)\mu \nu}^{\alpha \beta}W_{\left(\text{E}\right)\alpha \beta}^{\mu \nu}=W_{\left(\text{E}\right)kl}^{ij}W_{\left(\text{E}\right)ij}^{kl}+4 E_{\left(\text{E}\right)j}^{i}E_{\left(\text{E}\right)i}^{j}+4W_{\left(\text{E}\right)jk}^{iz}W_{\left(\text{E}\right)iz}^{jk}\,.
\label{weylsquareddecomp}
\end{equation}
Considering Eq.~\eqref{weylfalloff}, it is clear that the asymptotic behavior of the last two terms is either of the same order as the Pfaffian or it falls-off even faster. Thus, the only divergent contribution is coming from the first term as
\begin{equation}
\left\vert W_{\left(\text{E}\right)} \right\vert ^{2}=W_{\left(\text{E}\right)kl}^{ij}W_{\left(\text{E}\right)ij}^{kl}+\mathcal{O}\left(  z^{6}\right) \sim \mathcal{O}\left(  z^{4}\right) \,.
\label{weylsquaredbulkdiverge}
\end{equation}
Furthermore, taking into account the Gauss-Codazzi decomposition of the Riemann tensor, we obtain
\begin{align}
W_{\left(\text{E}\right)kl}^{ij}  &  =\mathcal{R}_{kl}^{ij}\left(h\right)-K_{k}^{i}K_{l}^{j}+K_{l}^{i}K_{k}^{j}+\frac{1}{\ell
^{2}}\delta_{kl}^{ij} \notag \\
& =\mathcal{R}_{kl}^{ij}-\delta_{ k}^{i}\mathcal{S}
_{l}^{j} +\delta_{ l}^{i}\mathcal{S}
_{k}^{j}-\delta_{ l}^{j}\mathcal{S}
_{k}^{i}+\delta_{ k}^{j}\mathcal{S}
_{l}^{i}+\mathcal{O}\left( z^{4}\right) \notag \\
&  =\mathcal{W}_{kl}^{ij} \left(h\right)+\mathcal{O}\left( z^{4}\right) \,,
\end{align}
where the expansion of the extrinsic curvature
\begin{equation}
K_{k}^{i}=\frac{1}{\ell}\delta_{k}^{i}%
+\ell\mathcal{S}_{k}^{i} \left(h\right)+\mathcal{O}\left( z^{2}\right)\,,
\end{equation}
has been taken and $\mathcal{R}_{kl}^{ij}$ is the Riemann tensor of the codimension-1 radial hypersurface.
Thus, the bulk Weyl tensor with intrinsic indices of the radial hypersurface matches to the leading order the intrinsic Weyl tensor. As a consequence, we obtain for the first term of Eq.~\eqref{weylsquareddecomp} that
\begin{equation}
\left\vert  W_{\left(\text{E}\right)}\right\vert ^{2}= W_{\left(\text{E}\right)kl}^{ij}W_{\left(\text{E}\right)ij}^{kl}+ \mathcal{O}\left(
z^{6}\right) =z^{2}\mathcal{W}_{kl}^{ij} \left(g_{(0)}\right) \mathcal{W}_{ij}^{kl} \left(g_{(0)}\right)+\mathcal{O}\left(
z^{6}\right)  \,.
\end{equation}
Hence, we realize that it is the fall-off of the squared Weyl tensor in the radial hypersurface that determines the divergence structure of the topologically renormalized action. Indeed, its integral, after performing the radial integration and inverting the FG series, is given by
\begin{equation}
{\displaystyle\int\limits_{M_{6}}}
\diff{^6x}\sqrt{|g|}\,\left\vert W_{\left(\text{E}\right)} \right\vert ^{2} =
{\displaystyle\int\limits_{\partial M_{6}}}
\diff{}^{5}x\sqrt{|h|}\left\vert \mathcal{W} \left(h\right)\right\vert ^{2}+\text{finite} \,.
\end{equation}
For ACF manifolds, the leading-order contribution falls-off fast enough, such that the action in Eq.~\eqref{EH2} is finite. This term vanishes trivially for spacetimes with conformally flat radial slicings. In all the other cases, the last equation diverges, and the topologically renormalized Einstein-AdS action should be supplemented by additional terms such that
\begin{align}\label{IEHren}
I_{\rm EH}^{\rm (ren)} = \kappa\ell^4\int_{\mathcal{M}_6}\diff{^6x}\sqrt{|g|}\left[P_6\left(W_{\left(\text{E}\right)}\right) + \frac{1}{48}\Box\left(W^{\nu\lambda}_{\left(\text{E}\right)\rho\sigma}W^{\rho\sigma}_{\left(\text{E}\right)\nu\lambda} \right) \right]+\tilde{\chi}_6\,.
\end{align}
In the next section, we  determine the counterterm that appears naturally from the embedding of Einstein gravity into CG in six dimensions, introducing the concept of conformal renormalization. 

\section{Einstein gravity from six-dimensional conformal gravity}
\label{ConformalRenormalization}

The definition of Conformal Gravity in even dimension depends on the construction and classification of the local conformal invariants of the corresponding dimension. In 4D the unique conformally covariant scalar is the Weyl tensor squared. Even though one determines trivial conformal invariants constructed out of contractions of the Weyl in higher dimensions, there are non-trivial cohomologies that contribute such that additional conformal scalars arise. In particular, in six dimensions these invariants obtain the form
\begin{subequations}\label{conformalinvariants}
    \begin{align}
    I_{\rm CG}^{(1)} &= \int_{\mathcal{M}_6}\diff{^6x}\sqrt{|g|} W_{\alpha\beta\mu\nu}W^{\alpha\rho\lambda\nu}W_{\rho}{}^{\beta\mu}{}_{\lambda} \,, \\
    I_{\rm CG}^{(2)} &=  \int_{\mathcal{M}_6}\diff{^6x}\sqrt{|g|} W_{\mu\nu\alpha\beta}W^{\alpha\beta\rho\lambda}W_{\rho\lambda}{}^{\mu\nu} \,, \\
    I_{\rm CG}^{(3)} &=  \int_{\mathcal{M}_6}\diff{^6x}\sqrt{|g|}\left[W_{\mu\rho\sigma\lambda}\left(\delta^\mu_\nu\Box + 4R^\mu_\nu - \frac{6}{5}R\delta^\mu_\nu \right)W^{\nu\rho\sigma\lambda} + \nabla_\mu V^\mu\right] \,, \\
    \notag
    V_\mu &= 4 R_{\mu}{}^{\lambda\rho\sigma}\nabla^\nu R_{\nu\lambda\rho\sigma} + 3R^{\nu\lambda\rho\sigma}\nabla_\mu R_{\nu\lambda\rho\sigma} - 5R^{\nu\lambda}\nabla_\mu R_{\nu\lambda} \\
    &\quad + \frac{1}{2}R\nabla_\mu R - R^\nu_\mu\nabla_\nu R + 2R^{\nu\lambda}\nabla_\nu R_{\lambda\mu} \,.
\end{align}
\end{subequations}
As a consequence, Conformal Gravity is a six-derivative metric theory defined as the arbitrary combinations of $I_{CG}^{\left(i\right)}$, namely
\begin{align}\label{ICG}
 I_{\rm CG} &=\alpha_{1} I_{\rm CG}^{(1)} + \alpha_{2} I_{\rm CG}^{(2)}  + \alpha_{3}I_{\rm CG}^{(3)} \,.
\end{align}
In Ref.~\cite{Lu:2013hx} it was shown that this theory admits Schwarzschild-AdS BH in its solutions space when choosing $\alpha_{1}=4 \alpha, \alpha_{2}=\alpha, \alpha_{3}=-\frac{1}{3} \alpha$.

One may provide an alternative derivation of this particular CG theory by applying reverse engineering from the Einstein-AdS action~\eqref{EH2} renormalized by the addition of a topological term. In the 4D case, for instance, the renormalized action can be thought of as a sector of CG~\cite{Anastasiou:2020mik}. \emph{Conformal covariantization} of Einstein-AdS amounts to embedding it in a conformally invariant theory, which in four dimensions is given by a Weyl-squared Lagrangian. This is achieved by replacing $W_{\left(\text{E}\right)} $ by the full Weyl tensor.

However, this strategy is not enough in the 6D case. Indeed, applying naively the previous rule on $P_6\left(W_{\rm (E)}\right)$, it will give rise to the Pfaffian of the Weyl tensor plus an additional four-derivative term proportional to Weyl-squared. This term should be supplemented by a curvature term, such that the entire expression reduces to $P_6\left(W_{\rm (E)}\right)$ when evaluated in Einstein-AdS spaces. This is given by the Schouten tensor, which becomes $S_{ \left(\text{E}\right)\nu}^{\mu} = -\frac{1}{2 \ell^2} \delta_{\nu}^{\mu}$ for Einstein spacetimes. In total, the six-derivative object that reduces to the Einstein-AdS action \eqref{EH} when evaluated for Einstein spaces is given by
\begin{equation}
\Lag_{6} = \delta _{\nu _{1}\ldots \nu _{6}}^{\mu _{1}\ldots \mu _{6}}W_{\mu _{1}\mu _{2}}^{\nu _{1}\nu _{2}}W_{\mu _{3}\mu _{4}}^{\nu _{3}\nu _{4}}W_{\mu _{5}\mu _{6}}^{\nu _{5}\nu _{6}} +12\delta _{\nu _{1}\ldots \nu _{5}}^{\mu _{1}\ldots \mu _{5}}W_{\mu _{1}\mu _{2}}^{\nu _{1}\nu _{2}}W_{\mu _{3}\mu _{4}}^{\nu _{3}\nu _{4}}S_{\mu _{5}}^{\nu _{5}} \,.
\label{L6}
\end{equation}
Clearly, this cannot describe 6D CG since the second term is not Weyl covariant. Performing the Weyl completion of this term (see Appendix~\ref{Weyl completion} for details) one arrives at the expression
\begin{align}\label{ICG2}
I_{\rm LPP} = \int_{\mathcal{M}_6}\diff{^6x}\sqrt{|g|}\,\Lag_{\rm LPP} + \int_{\partial\mathcal{M}_6}\diff{^5x}\sqrt{|h|}\,n_\mu J^\mu \,,
\end{align}
where
\begin{align}
\Lag_{\rm LPP} &= \alpha\left(\frac{1}{4!}\delta_{\mu_1\ldots\mu_6}^{\nu_1\ldots\nu_6}W^{\mu_1\mu_2}_{\nu_1\nu_2}W^{\mu_3\mu_4}_{\nu_3\nu_4}W^{\mu_5\mu_6}_{\nu_5\nu_6} + \frac{1}{2}\delta_{\mu_1\ldots\mu_5}^{\nu_1\ldots\nu_5}W^{\mu_1\mu_2}_{\nu_1\nu_2}W^{\mu_3\mu_4}_{\nu_3\nu_4}S^{\mu_5}_{\nu_5} + 8C^{\mu\nu\lambda}C_{\mu\nu\lambda}\right)\,, \\
\label{Jmu}
J^\mu &= \alpha\left(8 W^{\mu\kappa\lambda\nu}C_{\kappa\lambda\nu} - W^{\kappa\lambda}_{\nu\sigma}\nabla^\mu W^{\nu\sigma}_{\kappa\lambda}\right)\,.
\end{align}
As shown in Ref.~\cite{Anastasiou:2020mik}, this action is equal to the Lu-Pang-Pope CG~\cite{Lu:2013hx}. Indeed, starting from Eq.~\eqref{ICG2} and introducing the Weyl Bianchi identity
\begin{equation}
4I_{\rm CG}^{(1)}-I_{\rm CG}^{(2)}=W_{\mu\nu}^{\kappa\lambda}\nabla^{2}W_{\kappa\lambda}^{\mu\nu
}-8S_{\rho}^{\omega}W_{\nu\lambda}^{\rho\mu}W_{\omega\mu}^{\nu\lambda
}-2SW_{\nu\lambda}^{\rho\mu}W_{\rho\mu}^{\nu\lambda}+24C^{\mu\nu\lambda}%
C_{\mu\nu\lambda}+8\nabla_{\omega}\left(  W^{\omega\mu\nu\lambda}C_{\mu
\nu\lambda}\right)  \,,
\label{Bianchi}
\end{equation}
along with Eq.~\eqref{Pfaffian}, one ends up with the six-dimensional CG proposed in Ref.~\cite{Lu:2013hx} for $\alpha_{1}=4 \alpha, \alpha_{2}=\alpha, \alpha_{3}=-\frac{1}{3} \alpha$. The detailed analysis can be found in Appendix~\ref{AppB}. 

It is worthwhile to stress that, in the original work \cite{Lu:2013hx} it was shown that the theory admits the six-dimensional Schwarzschild-AdS black hole as a solution. As an extension of this result, in Ref.~\cite{Anastasiou:2020mik} it was shown that the whole set of Einstein spaces is a solution to LPP CG. Furthermore, the action principle can be written on a more convenient basis from a conformal geometry viewpoint (in terms of the Weyl, Schouten, and Cotton tensors),
what plays a key role when evaluating it on Einstein spaces.\footnote{The change of basis from the Lu-Pang-Pope form of Ref.~\cite{Lu:2013hx} to Eq.~\eqref{ICG2} introduces redundancies in the form of Bianchi identities, that do not modify the physical content of the theory. We thank Nicolas Boulanger for clarifying this issue.} 

The equations of motion for purely metric gravity theories that depend on first order covariant derivatives of the Riemann, i.e. of the form $\Lag \left( g_{\mu \nu}, R_{\mu \nu}^{\lambda \kappa},\nabla_{\alpha} R_{\mu \nu}^{\lambda \kappa}\right)$, can be written as~\cite{Araya:2021atx}
\begin{align}\notag
\mathcal{E}^\mu_\nu &= P^{\mu\lambda}_{\rho\sigma}R^{\rho\sigma}_{\nu\lambda}  - \frac{1}{2}\delta^\mu_\nu \Lag + 2\nabla_\lambda \nabla^\rho P^{\mu\lambda}_{\nu\rho} - 2 \nabla^\rho \nabla_\lambda \nabla_\sigma 
Q^{\mu\lambda\sigma}_{\nu\rho} + \frac{1}{2}Q^{\kappa\rho\mu}_{\sigma\tau}\nabla_\nu R^{\sigma\tau}_{\kappa\rho} + Q^{\mu\sigma\rho}_{\tau\kappa}\nabla_\rho R^{\tau\kappa}_{\nu\sigma} \\
\label{EOMCG}
&\quad - \nabla_\rho\left(2Q^{\tau\kappa\mu}_{\nu\sigma}R^{\sigma\rho}_{\tau\kappa} + 2Q^{\rho\sigma\mu}_{\tau\kappa}R^{\tau\kappa}_{\nu\sigma} + Q^{\mu\sigma\rho}_{\tau\kappa}R^{\tau\kappa}_{\nu\sigma}\right)\,,
\end{align}
 where the tensors considered are defined as
\begin{align}\label{Ptensor}
    P^{\mu\nu}_{\lambda_\rho} = \frac{\partial\Lag}{\partial R^{\lambda\rho}_{\mu\nu}} \;\;\;\;\; \mbox{and} \;\;\;\;\; 
    Q^{\mu\nu\sigma}_{\lambda\rho} = \frac{\partial\Lag}{\partial\nabla_\sigma R^{\lambda\rho}_{\mu\nu}} \,.
\end{align}
For the particular case of LPP Conformal Gravity, their explicit form is given by
\begin{align}\label{Ptensor}
    P^{\mu\nu}_{\lambda_\rho} &= \frac{\alpha}{8}\delta^{\mu\nu\mu_3\ldots\mu_6}_{\lambda\rho\nu_3\ldots\nu_6}W^{\nu_3\nu_4}_{\mu_3\mu_4}\left(W^{\nu_5\nu_6}_{\mu_5\mu_6} + 8S^{\nu_5}_{\mu_5}\delta^{\nu_6}_{\mu_6} \right) - 8\alpha\left(\Delta^\alpha_\beta \right)^{\mu\nu}_{\lambda\rho}\delta^{\beta\mu_2\mu_3\mu_4}_{\alpha\mu_2\mu_3\mu_4}W^{\nu_2\nu_3}_{\mu_2\mu_3}S^{\nu_4}_{\mu_4}\,, \\
    Q^{\mu\nu\sigma}_{\lambda\rho} &= 32\alpha\delta^\sigma_\beta\left(\Delta^\alpha_\gamma \right)^{\mu\nu}_{\lambda\rho}C_\alpha^{\ \beta\gamma}\,.
\end{align}
Here, $\left(\Delta^\alpha_\beta \right)^{\mu\nu}_{\lambda\rho}$ projects the components of the Riemann tensor onto the Schouten tensor, that is,  $S^\alpha_\beta = \left(\Delta^\alpha_\beta \right)^{\mu\nu}_{\lambda\rho}R^{\lambda\rho}_{\mu\nu}$, and it can be written explicitly as
\begin{align}
    \left(\Delta^\alpha_\beta \right)^{\mu\nu}_{\lambda\rho} = - \frac{1}{16}\left(\delta^{\alpha\mu\nu}_{\beta\lambda\rho} - \frac{4}{5}\delta^\alpha_\beta\delta^{\mu\nu}_{\lambda\rho} \right)\,.
\end{align}
The field equations for LPP gravity define a Weyl covariant, symmetric, traceless, conserved tensor $H^{\mu\nu}=\mathcal{E}^{\mu\nu}$~\cite{Lu:2013hx}.
In what follows, we show that any Einstein space is a solution to $H^{\mu\nu}=0$.  
In other words, this object appears as an \emph{obstruction} tensor in six dimensions, and naturally generalizes the properties of the Bach tensor in 4D. This provides support to the idea of embedding Einstein-AdS into CG at the level of the equations of motion and for obtaining the proper counterterms upon evaluating these spaces on the action. 

\subsection*{Einstein spaces as generic solutions of LPP CG in 6D}

For Einstein spaces, the Ricci tensor is $R_{\mu\nu}=-\frac{5}{\ell^2}\,g_{\mu\nu}$, the Schouten becomes $S^\mu_\nu = -\tfrac{1}{2\ell^2}\delta^\mu_\nu$ and, therefore, the Cotton tensor and $Q^{\mu\nu\sigma}_{\lambda\rho}$ vanish identically. Then, since the Weyl tensor is traceless, the last term of Eq.~\eqref{Ptensor} is zero as well. Furthermore, using Eq.~\eqref{WeylE}, the $P$-tensor evaluated on Einstein spaces becomes
\begin{align}\label{PtensorE}
    P^{\mu\nu}_{\lambda\rho}\big|_{\rm E} = \frac{\alpha}{8}\delta^{\mu\nu\mu_3\ldots\mu_6}_{\lambda\rho\nu_3\ldots\nu_6}\left(R^{\nu_3\nu_4}_{\mu_3\mu_4}R^{\nu_5\nu_6}_{\mu_5\mu_6} - \frac{1}{\ell^4}\delta^{\nu_3\nu_4}_{\mu_3\mu_4}\delta^{\nu_5\nu_6}_{\mu_5\mu_6}\right) \,.
\end{align}
Moreover, the Lagrangian takes the form $\Lag_{\rm{LPP}}|_{\rm E}=-24\alpha P_6(W_{\rm (E)})$, where $P_6(W_{\rm (E)})$ has been defined in Eq.~\eqref{P6WE} and it is equal to the parenthesis of Eq.~\eqref{EH}. Both simplifications make proving that $H^\mu_\nu \big|_{\rm E} =0$ considerably easy. First of all, by virtue of Bianchi identities, the third term in Eq.~\eqref{EOMCG} vanishes. Then, the field equations evaluated on Einstein spaces can be written as
\begin{align}
H^\mu_\nu \big|_E =& \frac{\alpha}{8} \left(\delta^{\rho\sigma\mu_3\ldots\mu_6}_{\nu\lambda\nu_3\ldots\nu_6} R^{\nu_3\nu_4}_{\mu_3\mu_4}R^{\nu_5\nu_6}_{\mu_5\mu_6} R^{\mu \lambda}_{\rho \sigma}-\frac{1}{6} \delta^{\mu} _{\nu} \delta_{\mu_1\ldots\mu_6}^{\nu_1\ldots\nu_6}R^{\mu_1\mu_2}_{\nu_1\nu_2}R^{\mu_3\mu_4}_{\nu_3\nu_4}R^{\mu_5\mu_6}_{\nu_5\nu_6}\right) \nonumber \\
&-\frac{\alpha}{8\ell^4} \delta^{\rho\sigma\mu_3\ldots\mu_6}_{\nu\lambda\nu_3\ldots\nu_6} \delta^{\nu_3\nu_4}_{\mu_3\mu_4} \delta^{\nu_5\nu_6}_{\mu_5\mu_6} R^{\mu \lambda}_{\rho \sigma} +\frac{12 \alpha}{\ell^4} \delta^{\mu} _{\nu} \left(R+\frac{20}{\ell^2}\right) \,.
\label{EOMeinstein}
\end{align}
The first line in the equation above comes from the variation of the Euler invariant $\mathcal{E}_6$ which, due to its topological nature, does not contribute to the field equations. Indeed, its variation produces something proportional to
\begin{equation}
\delta^{\mu}_{\nu} \delta^{\mu_1\ldots\mu_6}_{\nu_1\ldots\nu_6}R^{\nu_1\nu_2}_{\mu_1\mu_2} R^{\nu_3\nu_4}_{\mu_3\mu_4} R^{\nu_5\nu_6}_{\mu_5\mu_6}-6 \delta^{\mu_1\ldots\mu_5 \mu_6}_{\nu_1\ldots\nu_5 \nu }R^{\nu_1\nu_2}_{\mu_1\mu_2}R^{\nu_3\nu_4}_{\mu_3\mu_4} R^{\nu_5\mu}_{\mu_5\mu_6} = \delta^{\mu\mu_1\ldots\mu_6}_{\nu\nu_1\ldots\nu_6}R^{\nu_1\nu_2}_{\mu_1\mu_2}R^{\nu_3\nu_4}_{\mu_3\mu_4} R^{\nu_5\nu_6}_{\mu_5\mu_6} \,.
\end{equation}
The right-hand side in the last equation is identically zero, as the number of antisymmetrized indices exceeds the dimension of the manifold. Thus, the first line of Eq.~\eqref{EOMeinstein} vanishes identically. The rest of the field equations, constructed from the second term in Eq.~\eqref{EOMeinstein}, turn out to be proportional to the Einstein equation, namely, 
\begin{equation}
H^\mu_\nu \big|_E=-\frac{24}{\ell^2}\left(R^\mu_\nu -\frac{1}{2} R \delta^\mu_\nu -\frac{10}{\ell^2}\delta^\mu_\nu \right) =0\,.
\end{equation}
This argument proves, in full generality, that Einstein spaces are a sector of the solution space of LPP conformal gravity. Therefore, as the theory has Weyl symmetry, conformally-Einstein spaces are also a solution.

On the other hand, evaluating the action functional~\eqref{ICG2} in Einstein spaces and using Eq.~\eqref{WeylE}, we obtain
\begin{align}\label{ICGE}
    I_{\rm CG}^{\rm (E)} &= -24\,\alpha\int_{\mathcal{M}_6}\diff{^6x}\sqrt{|g|}\left[P_6 \left(W_{\left(\text{E}\right)}\right) + \frac{1}{48}\Box\left(W^{\nu\lambda}_{\left(\text{E}\right)\rho\sigma}W^{\rho\sigma}_{\left(\text{E}\right)\nu\lambda} \right) \right]\,.
\end{align}
Thus, fixing $\alpha=-\frac{\kappa \ell^4}{24}$, one arrives exactly to the renormalized action~\eqref{IEHren} including the extra counterterm that cancels the divergence generated by non-conformally flat hypersurfaces of constant bulk radius.\footnote{If one starts from the CG theory given directly by the linear combination of the conformal invariants of Eq.~\eqref{conformalinvariants}, one misses the topological number $\tilde{\chi}_6$ defined in Eq.~\eqref{chitilde}. However, this constant can be added as part of the definition of the CG theory without issue.} 
Indeed, as it was shown in the previous section, the Pfaffian part of $P_6$ is finite, whereas the Weyl-squared term contributes with IR divergences when $g_{\left(0\right)}$ is not conformally flat. This is canceled by the addition of the surface term seen in Eq.~\eqref{ICGE}, as it was proven in Ref.~\cite{Anastasiou:2020mik}. In this sense, embedding Einstein-AdS gravity into CG dictates exactly the series of counterterms in a closed form. This procedure is what we dubbed Conformal Renormalization.

\section{Topological black holes in 6D\label{sec:Schw}}

In this section, we compute explicitly the Euclidean on-shell action for topological black hole in six dimensions, that is, static black hole solutions with an arbitrary four-dimensional manifold as a transverse section. To this end, we consider the metric ansatz in Euclidean signature
\begin{align}\label{Schwansatz}
 \diff{s^2} = f(r)\diff{\tau^2} + \frac{\diff{r^2}}{f(r)} + r^2\diff{\Sigma^2_{(4)}}\,, \;\;\;\;\; \mbox{where} \;\;\;\;\; \diff{\Sigma^2_{(4)}} = \sigma_{mn}(x)\diff{x^m}\diff{x^n}
\end{align}
denotes the line element of the transverse section with $x^m$ being local coordinates on the codimension-2 hypersurface. Indeed, the Einstein equations imply that $\diff{\Sigma_{(4)}^2}$ has to be an Einstein space.

In $D=4$, every codimension-$2$ transverse section has a constant curvature and it is conformally flat. In higher dimensions, however, this is no longer true: one could have either conformally flat or non-conformally flat Einstein metrics, even without constant curvature. For instance, it is known that $\mathbb{S}^4$, $\mathbb{H}^4$, and $\mathbb{T}^4$ have constant curvature, i.e. $\mathcal{R}^{m_1m_2}_{n_1n_2}(\sigma)=\gamma\,\delta^{m_1m_2}_{n_1n_2}$ for $\gamma=\pm1,0$, respectively. Additionally, they are conformally flat, that is, $\mathcal{W}^{m_1m_2}_{n_1n_2}(\sigma)=0$  and, locally, they can be described by the line element
\begin{align}\label{S4T4H4}
    \diff{\Sigma_{(4)}^2} =  \frac{\delta_{mn}\diff{x^m}\diff{x}^n}{\left[1+\frac{\gamma}{4}\,\vec{x}\cdot\vec{x} \right]^2}\,,
\end{align}
where $\vec{x}\cdot\vec{x}=\delta_{mn}x^m\,x^n$. 
Nevertheless, one could consider other Einstein spaces as transverse sections, e.g. $\mathbb{S}^2\times\mathbb{S}^2$, $\mathbb{H}^2\times\mathbb{H}^2$, $\mathbb{CP}^2$, $\mathbb{CH}^2$, which are non-conformally flat, i.e. $\mathcal{W}^2(\sigma)\neq0$. Furthermore, the last two do not have constant curvature and they are self-dual with respect to the codimension-2 Levi-Civita tensor, i.e. $\mathcal{W}_{m_1m_2m_3m_4}(\sigma) = \tfrac{1}{2}\varepsilon_{m_1m_2n_1n_2}\mathcal{W}^{n_1n_2}_{m_3m_4}(\sigma)$. They are usually regarded as self-dual gravitational instantons in four-dimensional Einstein gravity with a nonvanishing cosmological constant~\cite{Eguchi:1980jx}. The line element of  $\mathbb{S}^2\times\mathbb{S}^2$ and $\mathbb{H}^2\times\mathbb{H}^2$ can be parametrized as
\begin{align}\label{S2S2H2H2}
    \diff{\Sigma^2_{(4)}} = \sum_{I=1}^2\frac{\diff{u_I^2}+\diff{v_I^2}}{\left[1+\frac{\gamma}{4}\left(u_I^2+v_I^2 \right) \right]^2}\,,
\end{align}
for $\gamma=\pm1$, respectively. On the other hand, the line element of the complex projective space, $\mathbb{CP}^2$, can be written as
\begin{align}\label{CPK}
 \diff{\Sigma_{(4)}^2} &= 6\bigg[\diff{\psi_2^2} + \sin^2\psi_2\cos^2\psi_2\left(\diff{\phi_2} + \sin^2\psi_1\diff{\phi_1} \right)^2 
 + \sin^2\psi_2\diff{\Omega^2} \bigg]\,,
 \end{align}
where $0\leq\psi_I\leq\pi/2$ and $0\leq\phi_I\leq 2\pi$ with $I=1,2$. The complex hyperbolic space $\mathbb{CH}^k$, on the other hand, can be obtained from a similar slicing, but replacing $\cos\psi_2\to\cosh\psi_2$ and $\sin\psi_2\to\sinh\psi_2$ where, in this case, $-\infty<\psi_I<\infty$ and $0\leq\phi_I\leq 2\pi$.

Replacing the ansatz in Eq.~\eqref{Schwansatz} with transverse sections~\eqref{S4T4H4}--\eqref{CPK} into the Einstein field equations~\eqref{EinsteinEq}, one finds that the metric function that solves the system is
\begin{align}\label{fsol}
    f(r) = k - \frac{4\pi mG}{r^3\,\Sigma_{(4)}} + \frac{r^2}{\ell^2}\,,
\end{align}
where $m$ is an integration constant associated to the mass (see section~\ref{NW}) and $\Sigma_{(4)} = \int\diff{\Sigma_{(4)}}$ is the volume of the transverse section. Additionally, for $\mathbb{S}^4$, $\mathbb{H}^4$, and $\mathbb{T}^4$, we have that $k=\pm1,0$, respectively. For $\mathbb{CP}^2$ and $\mathbb{S}^2\times\mathbb{S}^2$, on the other hand, one finds that $k=1/3$, while for $\mathbb{CH}^2$ and $\mathbb{H}^2\times\mathbb{H}^2$ its value is $k=-1/3$. The horizon of this solution is defined by the largest positive root of the polynomial $f(r_h)=0$. The absence of conical singularities at the horizon is guaranteed if the Euclidean-time coordinate is identified as $\tau\sim\tau+\beta$, where
\begin{align}\label{betasol}
    \beta = \frac{4\pi r_h\ell^2}{3k\ell^2+5r_h^2}\,.
\end{align}
This value represents the inverse of the Hawking temperature of the topological Schwarzschild-AdS black hole. Then, the space is completely regular provided that $0\leq\tau\leq\beta$ and $r_h\leq r<\infty$, with the proper range of the codimension-2 boundary coordinates.

The global properties of this solution are labeled by the Euler characteristic and the Euclidean on-shell action. The former can be obtained directly from the Euler theorem~\eqref{Eulertheorem}. Indeed, the K\"unneth theorem for direct product spaces implies that 
\begin{align}
\chi(\mathcal{M}_6)=\chi(\mathbb{D}_2\times\Sigma)=\chi(\mathbb{D}_2)\times\chi(\Sigma)=1\times\chi(\Sigma)\,,    
\end{align}
where $\mathbb{D}_2$ is topologically the disk in the $\tau-r$ Euclidean section and $\Sigma$ is the transverse section which has the topology of the black hole horizon. Direct computation of the Euler theorem~\eqref{Eulertheorem} yields
\begin{align}\label{EulerSchw}
    \chi(\mathcal{M}_6) = \frac{3k^2\Sigma_{(4)}}{4\pi^2}\;\;\;\;\; \mbox{and} \;\;\;\;\; \chi(\mathcal{M}_6) = \frac{9k^2\Sigma_{(4)}}{4\pi^2}\,,
\end{align}
for conformally and non-conformally flat codimension-1 hypersurfaces of constant radius, respectively.

The Euclidean on-shell action is another relevant global quantity since, to first order in the saddle-point approximation, it allows one to compute the partition function $\mathcal{Z}$ through the relation $\ln\mathcal{Z}\approx-I_E$. All the thermodynamic properties of the solution can be obtained through standard statistical mechanics relations from the partition function. Considering a radial foliation [see Eq.~\eqref{GaussNormal}], one can check that the codimension-1 hypersurfaces with topology of $\mathbb{S}^1\times\Sigma$ are conformally flat if the codimension-2 transverse section is conformally flat, e.g. $\Sigma=\mathbb{S}^4,\mathbb{H}^4,\mathbb{T}^4$. Even more, the Einstein-AdS action augmented by the Euler density in Eq.~\eqref{EH2} provides a renormalized Euclidean on-shell action given by
\begin{align}\label{IErenSchwNew}
    -I_E &= \frac{k^2\ell^4\Sigma_{(4)}}{4G} - \tilde{\chi}_6 + \frac{\beta r_h^3\Sigma_{(4)}}{16\pi G\ell^2}\left(k\ell^2-r_h^2 \right)\,.
\end{align}
The value of $\tilde{\chi}_6$ can be computed directly from its definition in Eq.~\eqref{chitilde}, using the values of $\chi(\mathcal{M}_6)$ and $\eta$ given in Eqs.~\eqref{EulerSchw} and~\eqref{eta}, respectively. They give $\tilde{\chi}_6=\frac{k^2\ell^4\Sigma_{(4)}}{4G}$. Therefore, the first two terms on the right-hand side of Eq.~\eqref{IErenSchwNew} cancel each other and the renormalized Euclidean on-shell action for the topological black hole in six dimensions with conformally flat hypersurfaces of constant $r$ is 
\begin{align}\label{IErenfinal}
  -I_E =  \frac{\beta r_h^3\Sigma_{(4)}}{16\pi G\ell^2}\left(k\ell^2-r_h^2 \right)\,.
\end{align}
Although this result holds for any conformally flat codimension-1 hypersurface~\cite{Anastasiou:2020zwc}, this is not the case in the presence of non-conformally flat slices, e.g. when they have the topology of $\mathbb{S}^1\times\Sigma$, with $\Sigma=\mathbb{S}^2\times\mathbb{S}^2, \mathbb{H}^2\times\mathbb{H}^2, \mathbb{CP}^2,\mathbb{CH}^2$. In this case, $\left \vert \mathcal{W} \left(h\right)\right \vert ^2=r^{-4}\left \vert \mathcal{W} \left(\sigma\right) \right \vert  ^2$, what switches on the divergent part of the bulk Weyl tensor squared in Eq.~\eqref{weylsquaredbulkdiverge}, that spoils the ACF condition. Indeed, evaluating the Euclidean on-shell action~\eqref{EH2} for the solution~\eqref{fsol} with a non-conformally flat transverse section leads to a divergent result, that is, 
\begin{align}
 -I_E = -\kappa\ell^2\int_{\mathcal{M}_6}\diff{^6x}\sqrt{|g|}P_6(W_{\rm (E)}) = -4\Sigma_{(4)}\kappa k^2\ell^2\beta^2\, r\big|_{r=r_h}^{r=\infty} + \mbox{finite terms}\,.
\end{align}
Remarkably, by evaluating the solution in the action~\eqref{IEHren} which includes the counterterms that arise from CG [cf. Eq.~\eqref{ICGE}], we obtain the renormalized Euclidean on-shell action
\begin{align}\label{IErenSchw0}
 -I_{\rm EH}^{\rm (ren)} &=
\frac{3k^2\ell^4\Sigma_{(4)}}{4G} - \tilde{\chi}_6 + \frac{\beta r_h^3\Sigma_{(4)}}{16\pi G\ell^2}\left(k\ell^2-r_h^2 \right)\,.
\end{align}
In this case, the value of the topological number is $\tilde{\chi}_6=\frac{3k^2\ell^4\Sigma_{(4)}}{4G}$. Therefore, the action for topological BHs with generic transverse sections is that given in Eq.~\eqref{IErenfinal}, including the cases with $k=\pm1,0$ and $k=\pm1/3$.

Instead of Conformal Renormalization, one could have considered the holographic renormalization prescription~\cite{deHaro:2000vlm,Henningson:1998gx,Skenderis:2002wp,Papadimitriou:2004ap}, which consists of adding intrinsic boundary terms to the bulk action such that its finiteness and a well-posed variational principle are guaranteed. In six dimensions, the action functional is given by~\cite{Balasubramanian:1999re,Emparan:1999pm,deHaro:2000vlm,Skenderis:2002wp}
\begin{align}\notag
    I_{\rm HR} &= \kappa\int_{\mathcal{M}_6}\diff{^6x}\sqrt{|g|}\,\left(R+\frac{20}{\ell^2} \right) + 2\kappa\int_{\partial\mathcal{M}_6}\diff{^5x}\sqrt{|h|}\,K   \\
\label{IHR}    
    & + 2\kappa\int_{\partial\mathcal{M}_6}\diff{^5x}\sqrt{|h|}\left(\frac{4}{\ell} + \frac{\ell}{6} + \frac{\ell^3}{18}\left[\mathcal{R}_{ij}\mathcal{R}^{ij} - \frac{5}{16}\mathcal{R}^2 \right]+\ldots \right)\,.
\end{align}
Evaluating the action~\eqref{IHR} on the 
topological AdS black hole solution, we obtain exactly the same value as that given in Eq.~\eqref{IErenfinal}. This result provides a concrete example of the equivalence between the holographic and conformal renormalization schemes.

\section{Conserved charges and thermodynamics\label{NW}}

In order to compute the conserved charges, we adopt the Noether-Wald formalism~\cite{Iyer:1994ys}. In particular, we use the Noether prepotential derived in Ref.~\cite{Anastasiou:2021tlv} for six-dimensional conformal gravity, that is,
\begin{align}
    q^{\mu\nu} = - \left(P^{\mu\nu}_{\lambda\rho}\nabla^\lambda\xi^\rho + 2\xi^\lambda\nabla^\rho P^{\mu\nu}_{\lambda\rho} \right)=-q^{\nu\mu}\,,
\end{align}
where $P^{\mu\nu}_{\lambda\rho}$ is defined in Eq.~\eqref{Ptensor} and it is assumed to be evaluated on the Einstein sector of the space of solutions [see Eq.~\eqref{PtensorE}]. The Noether-Wald charges associated with the Killing vector field that generates the Euclidean-time symmetry, $\xi=\partial_\tau$, are the mass and Wald entropy given by
\begin{align}
     M &= \int_\infty\left(q^{\mu\nu} - 2\xi^{[\mu}J^{\nu]} \right)\diff{\Sigma}_{\mu\nu}\,, \\
    S &= \beta_\tau\int_{\mathcal{H}}q^{\mu\nu}\diff{\Sigma}_{\mu\nu}\,,
\end{align}
respectively, where $\beta_\tau$ is the period of the Euclidean time that equals the inverse of the Hawking temperature [see Eq.~\eqref{betasol}], $J^\mu$ is defined in Eq.~\eqref{Jmu}, and $\diff{\Sigma_{\mu\nu}}$ denotes the volume element of a codimension 2 hypersurface. 

For the topological BHs of Eq.~\eqref{fsol}, we obtain that the mass and entropy are 
\begin{align}
    M = m \;\;\;\;\; \mbox{and} \;\;\;\;\; S = \frac{r_h^4\Sigma_{(4)}}{4G} - \tilde{\chi}_6\,,
\end{align}
respectively. Notice that the entropy follows the standard one-quarter area law plus a constant related to the Euler characteristic of the horizon, which appears here because the Noether-Wald formalism is not sensitive to constant numbers present in the action, as it has a variational origin. The latter does not affect the first law of thermodynamics, i.e.,
\begin{align}
    \delta M = T\delta S \,.
\end{align}
Indeed, one could eliminate this constant by using the Euler theorem~\eqref{Eulertheorem} to replace the integral of the Euler density $\mathcal{E}_6$ in terms of the Chern form $B_5$, matching exactly the result obtained via holographic renormalization.  

As we mentioned in section~\ref{sec:Schw}, the Euclidean on-shell action~\eqref{IErenfinal} is a relevant global quantity for obtaining the thermodynamic quantities of the topological BH to first order in the saddle-point approximation. From this result, one can obtain the mass, entropy, and free energy through
\begin{align}
M &= \frac{\partial I_{\rm EH}^{\rm (ren)}}{\partial\beta} = m\,, \\
S &= \beta \frac{\partial I_{\rm EH}^{\rm (ren)}}{\partial\beta} - I_{\rm EH}^{\rm (ren)} = \frac{r_h^4\Sigma_{(4)}}{4G} \,,\\
F &= \beta^{-1}I_{\rm EH}^{\rm (ren)} = M - TS\,,
\end{align}
respectively. These thermodynamic quantities coincide, up to the Euler characteristic, with the ones obtained through the Noether-Wald formalism and they satisfy the first law of thermodynamics. 

To see whether the system develops a phase transition, first, we notice that the radius of the black hole horizon in terms of the Hawking temperature is given by
\begin{align}
    r_h^{(\pm)}(T) &= \frac{\ell}{5}\left(2\pi\ell T \pm \sqrt{4\pi^2\ell^2 T^2 - 15k} \right)\,.
\end{align}
The branches $r_h^{(+)}$ and $r_h^{(-)}$ are usually dubbed the large and small black holes, respectively. If $k\geq0$, the cosmic censorship conjecture demands that $4\pi^2\ell^2T^2\geq15k$. The minimum temperature of the system, $T_0$, is achieved when this inequality is saturated, i.e.
\begin{align}
T_0=\frac{\sqrt{15k}}{2\pi\ell} \,.
\end{align}
The case $k=1$ is the only one that is continuously connected to global AdS$_6$ in the limit when $m\to0$. The latter can be used as a ground state whose free energy is always zero. One can see that the free energy for the small black hole, i.e. $r_h=r_h^{(-)}(T)$, with $\Sigma=\mathbb{S}^4$, is monotonically increasing for $T\geq T_0$. However, the free energy of the large black hole, i.e. $r_h=r_h^{(+)}(T)$, becomes negative for $T>T_c$, where the critical temperature for $k=1$ is 
\begin{align}\label{Tcritica}
    T_c = \frac{3\sqrt{3}}{2\pi\ell}\,.
\end{align}
Therefore, we conclude that there is a phase transition between AdS$_6$ and the large black hole for temperatures higher than the critical one given in Eq.~\eqref{Tcritica}.

\section{Discussion}
We considered the renormalization of Einstein-AdS$_6$ gravity by embedding the theory into the Lu, Pang and Pope Conformal Gravity~\cite{Lu:2013hx}; i.e. the unique combination of conformal invariants that admits Einstein solutions. This embedding induces the surface term that cancels the bulk divergences for AAdS spaces and is what we refer to as Conformal Renormalization. In particular, we studied solutions corresponding to six-dimensional Einstein-AdS manifolds whose hypersurfaces in the radial foliation along the holographic coordinate are not conformally flat. In this case, the Kounterterm renormalization procedure, reviewed in section~\ref{TopologicalRenormalization}, fails to cancel all divergences, as there is a remaining one in the action that depends on the bulk Weyl tensor squared. However, the Conformal Renormalization procedure
reproduces the terms dictated by the Kounterterms scheme, and it also includes the required term to cancel the remaining divergence, being contained in the boundary term present in the $I_{\rm CG}^{(3)}$ conformal invariant as shown in Eq.~\eqref{conformalinvariants}. 

As discussed in section~\ref{ConformalRenormalization}, in order to obtain the conformally renormalized action, one simply evaluates the CG action of Eq.~\eqref{ICG2} for Einstein spacetimes, namely, considering the vanishing of the Cotton tensor, and the restricted forms of the Schouten and the Weyl tensors~\eqref{WeylE}. The action thus obtained, is entirely written in terms of $W^{\alpha \beta}_{\left(\text{E}\right)\mu \nu}$ and a boundary term which depends on a covariant derivative thereof, as shown in Eq.~\eqref{ICGE}.

By direct evaluation of the Euclidean on-shell action, when considering manifolds with non-Weyl-flat transverse sections, the Conformal Renormalization is explicitly verified and compared to the result obtained with the standard holographic renormalization prescription, finding agreement between the two methods. In particular, in section~\ref{sec:Schw}, we consider the case of topological-AdS BHs with arbitrary transverse sections. This class of solutions contains manifolds with either conformally flat or non-conformally flat constant-r hypersurfaces, showing that the range of validity of our prescription is generic.

The examples considered in this work constitute a non-trivial test of the Conformal Renormalization scheme, which is interesting because it relies on bulk Weyl symmetry in order to fix the counterterms required to render the on-shell action finite. Furthermore, the resulting action is consistent with a Dirichlet variational principle for the holographic sources~\cite{Anastasiou:2020mik}. The finiteness and the variational principle are both requirements in order to use the action as the generating functional for correlators in the dual CFT, in the context of the AdS/CFT correspondence. Currently, the Conformal Renormalization scheme is being explored in the context of eight-dimensional Einstein-AdS gravity, four-dimensional scalar-tensor theories of gravity, renormalization of codimension-2 energy functionals in the AdS/CFT correspondence, among others. Exciting developments can be expected in the future. 

\begin{acknowledgments}
We thank Gastón Giribet for comments and discussions. The work of GA is funded by ANID, Convocatoria Nacional Subvenci\'on a Instalaci\'on en la Academia Convocatoria A\~no 2021, Folio SA77210007. IJA is supported by ANID FONDECYT grants No.~11230419 and~1231133, and by ANID Becas Chile grant No.~74220042. IJA also acknowledges funding by ANID, REC Convocatoria Nacional Subvenci\'on a Instalaci\'on en la Academia Convocatoria A\~no 2020, Folio PAI77200097. IJA is grateful to Andrei Parnachev and the School of Mathematics at Trinity College Dublin for their hospitality. CC is partially supported by Agencia Nacional de Investigaci\'{o}n y Desarrollo (ANID) through FONDECYT grants No~11200025, 1230112, and~1210500. The work of RO has been funded by the FONDECYT Regular Grants 1230492 and 1231779, and ANILLO Grant ANID/ACT210100. 
\end{acknowledgments}

\appendix

\section{Weyl completion of $P_{6}$}
\label{Weyl completion}

We are interested in determining the Weyl completion of the scalar density constructed out of $\Lag_6$ in Eq.~\eqref{L6}. To this end, we consider the infinitesimal Weyl transformation of the metric as
\begin{equation}\delta _{\sigma }g_{\alpha \beta } =2\sigma \left(x\right)g_{\alpha \beta } \,.
\label{metric}
\end{equation}
Under this rescaling, the Ricci scalar and Schouten tensor transform as
\begin{gather}
\delta _{\sigma }R = -2\sigma R -2\left (D -1\right ) \square \sigma  \,, \label{Ricci} \\
\delta _{\sigma }S_{\mu \nu } = - \nabla _{\mu } \nabla _{\nu }\sigma  \,, \label{Schouten}
\end{gather}
respectively. The Weyl tensor with two upper and two lower indices transforms homogeneously under Weyl rescalings, that is,
\begin{equation}
\delta _{\sigma }W_{\mu \nu }^{\kappa \lambda} =-2 \sigma \left(x\right) W_{\mu \nu }^{\kappa \lambda}\,,
\label{Weyl}
\end{equation}
while the Cotton and Bach tensors transform as
\begin{align}
\delta _{\sigma }C_{\mu \nu \lambda } &= -W_{\kappa \mu \nu \lambda } \nabla ^{\kappa }\sigma  \,, \label{Cotton} \\
\delta _{\sigma }B_{\mu \nu } &= -2\sigma B_{\mu \nu } +\left (D -4\right )\left (C_{\mu \nu \lambda } +C_{\nu \mu \lambda }\right ) \nabla ^{\lambda }\sigma  \,, \label{Bach}
\end{align}
respectively. Different variants of Eq.~\eqref{Cotton} can be provided, for instance,
\begin{equation}
\delta _{\sigma }C_{\ \nu \lambda }^{\mu }=W_{\nu \lambda }^{\mu \rho } \nabla _{\rho }\sigma  -2\sigma C_{\ \nu \lambda }^{\mu }  \;\;\;\;\; \mbox{and} \;\;\;\;\; \quad \delta _{\sigma }C_{\nu }^{\ \mu \rho } =W_{\nu \lambda }^{\mu \rho } \nabla ^{\lambda }\sigma -4\sigma C_{\nu }^{\ \mu \rho } \,.
\label{cotton2}
\end{equation}
On the other hand, the Weyl transformation of the divergence of a vector field $A_{\mu }$ in six dimensions is given by
\begin{align}
\delta _{\sigma }\left (\sqrt{|g|}\, \nabla ^{\mu }A_{\mu }\right ) 
&=\sqrt{|g|} \nabla ^{\mu }\left (4\sigma A_{\mu } +\delta _{\sigma }A_{\mu }\right ) \,.
\label{weyldivA}
\end{align}
Based on these expressions, we analyze each one of the terms of Eq.~\eqref{L6} under infinitesimal Weyl transformations. For the first term, we obtain
\begin{align}\notag
\delta _{\sigma }\left (\sqrt{|g|}\delta _{\nu _{1}\ldots \nu _{6}}^{\mu _{1}\ldots \mu _{6}}W_{\mu _{1}\mu _{2}}^{\nu _{1}\nu _{2}}W_{\mu _{3}\mu _{4}}^{\nu _{3}\nu _{4}}W_{\mu _{5}\mu _{6}}^{\nu _{5}\nu _{6}}\right ) =\sqrt{|g|} \delta _{\nu _{1}\ldots \nu _{6}}^{\mu _{1}\ldots \mu _{6}}\left[\frac{1}{2}W_{\mu _{1}\mu _{2}}^{\nu _{1}\nu _{2}}W_{\mu _{3}\mu _{4}}^{\nu _{3}\nu _{4}}W_{\mu _{5}\mu _{6}}^{\nu _{5}\nu _{6}}\left (g^{ \mu \nu}\delta _{\sigma }g_{\mu \nu}\right ) \right. \\ \left.+3W_{\mu _{1}\mu _{2}}^{\nu _{1}\nu _{2}}W_{\mu _{3}\mu _{4}}^{\nu _{3}\nu _{4}}\delta _{\sigma }W_{\mu _{5}\mu _{6}}^{\nu _{5}\nu _{6}} \right] = \sqrt{|g|} \delta _{\nu _{1}\ldots \nu _{6}}^{\mu _{1}\ldots \mu _{6}}\left (6\sigma W_{\mu _{1}\mu _{2}}^{\nu _{1}\nu _{2}}W_{\mu _{3}\mu _{4}}^{\nu _{3}\nu _{4}}W_{\mu _{5}\mu _{6}}^{\nu _{5}\nu _{6}} -6\sigma W_{\mu _{1}\mu _{2}}^{\nu _{1}\nu _{2}}W_{\mu _{3}\mu _{4}}^{\nu _{3}\nu _{4}}W_{\mu _{5}\mu _{6}}^{\nu _{5}\nu _{6}}\right )  = 0 \,.
\end{align}
As expected, this contribution is Weyl invariant in 6D. For the second term of Eq.~\eqref{L6}, taking into account Eqs.~\eqref{Schouten} and~\eqref{Weyl}, one obtains that
\begin{align}
\delta _{\sigma }\left (\sqrt{|g|}\delta _{\nu _{1}\ldots \nu _{5}}^{\mu _{1}\ldots \mu _{5}}W_{\mu _{1}\mu _{2}}^{\nu _{1}\nu _{2}}W_{\mu _{3}\mu _{4}}^{\nu _{3}\nu _{4}}S_{\mu _{5}}^{\nu _{5}}\right ) 
&= -\sqrt{|g|}\delta _{\nu _{1}\ldots \nu _{5}}^{\mu _{1}\ldots \mu _{5}}W_{\mu _{1}\mu _{2}}^{\nu _{1}\nu _{2}}W_{\mu _{3}\mu _{4}}^{\nu _{3}\nu _{4}} \nabla ^{\nu _{5}} \nabla _{\mu _{5}}\sigma  \,.
\end{align}
As a consequence, the scalar density constructed out of $\Lag_6$ is not Weyl invariant, since
\begin{equation}
\delta _{\sigma }\left (\sqrt{|g|}\,\Lag_{6}\right ) = -12\sqrt{|g|}\,\delta _{\nu _{1}\ldots \nu _{5}}^{\mu _{1}\ldots \mu _{5}}W_{\mu _{1}\mu _{2}}^{\nu _{1}\nu _{2}}W_{\mu _{3}\mu _{4}}^{\nu _{3}\nu _{4}} \nabla ^{\nu _{5}} \nabla _{\mu _{5}}\sigma \,.
\label{L6nonhomog}
\end{equation}
In order to render this quantity Weyl invariant and to make contact with CG, we need to perform its Weyl completion for an arbitrary scaling function $\sigma \left(x\right)$. In the calculations that will be performed below, we consider the decomposition of the Weyl tensor given by the relation
\begin{equation}
W_{\mu \nu }^{\alpha \beta } =R_{\mu \nu }^{\alpha \beta } -4S_{[\mu }^{[\alpha }\delta _{\nu ]}^{\beta ]} \,.
\end{equation}
As a first step, we integrate by parts once the Eq.~\eqref{L6nonhomog}, and taking into account the definition of the Cotton tensor in terms of the Schouten~\eqref{BachCottonSchouten} along with Eq.~\eqref{cotton2}, we obtain
\begin{align}
\delta _{\sigma }\left (\sqrt{|g|}\Lag_{6}\right ) 
&= 48\sqrt{|g|}\left[\nabla ^{\mu }\left (8\sigma W_{\nu \mu }^{\alpha \beta }C_{\alpha \beta }^{\nu }+W_{\nu \mu }^{\alpha \beta }\delta _{\sigma }C_{\alpha \beta }^{\nu } - W_{\lambda \nu }^{\alpha \beta }W_{\alpha \beta }^{\lambda \nu } \nabla _{\mu }\sigma\right ) - 8W_{\mu \nu }^{\alpha \beta }C_{\alpha }^{\mu \nu } \nabla _{\beta }\sigma \right].
\label{deltaL61}
\end{align}
Our objective is to determine the terms that will compensate the right-hand side of the last expression. Their form is suggestive and helpful for finding the terms that will render the action conformally invariant. Our starting point is the Laplacian of the Weyl squared term. Based on Eq.~\eqref{weyldivA}, the latter scales as 
\begin{align}
\delta _{\sigma }\left [\sqrt{|g|} \,\Box\left(W_{\kappa \nu }^{\alpha \beta }W_{\alpha \beta }^{\kappa \nu }\right)\right ] 
&= -8\sqrt{|g|} \nabla ^{\mu }\left (W_{\kappa \nu }^{\alpha \beta }W_{\alpha \beta }^{\kappa \nu } \nabla _{\mu }\sigma \right ) \,,\label{weylvarboxweylsquared}
\end{align}
which allows us to write the third term of the right-hand side of Eq.~\eqref{deltaL61} as the Weyl variation of the Laplacian of the Weyl squared term.
On the other hand, the infinitesimal Weyl transformation of the density constructed out of the divergence of contractions of the Weyl with the Cotton gives 
\begin{gather}
\delta _{\sigma }\left [\sqrt{|g|} \nabla ^{\mu }\left (W_{\nu \mu }^{\alpha \beta }C_{\ \alpha \beta }^{\nu }\right )\right ] 
=\sqrt{|g|} \nabla ^{\mu }\left (2\sigma W_{\nu \mu }^{\alpha \beta }C_{\ \alpha \beta }^{\nu } +W_{\nu \mu }^{\alpha \beta }\delta _{\sigma }C_{\ \alpha \beta }^{\nu }\right ) \,,
\label{welvardivweylcotton}
\end{gather}
where Eq.~\eqref{cotton2} has been taken into account. Then, it is clear that the second term of the right-hand side of Eq.~\eqref{deltaL61} can be rewritten in a more convenient way by using Eq.~\eqref{welvardivweylcotton}.
Additionally, considering the Weyl variation of Cotton squared, one obtains
\begin{gather}
\delta _{\sigma }\left (\sqrt{|g|}C_{\mu\nu\lambda}C^{\mu \nu\lambda }\right ) 
=2\sqrt{|g|}C_{\lambda }^{\ \mu \nu }\left (\delta _{\sigma }C_{\ \mu \nu }^{\lambda } +2\sigma C_{\ \mu \nu }^{\lambda }\right ) \,,
\label{weylvarcottonsquared}
\end{gather}
due to Eq.~\eqref{cotton2}.
Finally, replacing Eqs.~\eqref{weylvarboxweylsquared},~\eqref{welvardivweylcotton} and~\eqref{weylvarcottonsquared} into Eq.~\eqref{deltaL61}, we get
\begin{equation}
\delta _{\sigma }\left (\sqrt{|g|}\Lag_{6}\right )
= -24\delta _{\sigma }\left\{\sqrt{|g|} \left[8C_{\mu\nu\lambda}C^{\mu \nu\lambda } + \nabla ^{\mu }(8W_{\mu \nu }^{\alpha \beta }C_{\ \alpha \beta }^{\nu } -W_{\kappa \nu }^{\alpha \beta } \nabla _{\mu }W_{\alpha \beta }^{\kappa \nu })\right]\right\} \,.
\end{equation}
As a consequence, the action defined by
\begin{equation}
I_{\rm LPP} =\int_{\mathcal{M}_6}\diff{^6x}\sqrt{|g|}\left [\frac{1}{24}\Lag_{6} +8C_{\mu \nu\lambda }C^{\mu \nu \lambda } + \nabla ^{\mu }(8W_{\mu \nu }^{\alpha \beta }C_{\ \alpha \beta }^{\nu } -W_{\kappa \nu }^{\alpha \beta } \nabla _{\mu }W_{\alpha \beta }^{\kappa \nu })\right ] \,
\end{equation}
is Weyl invariant, i.e., $\delta _{\sigma }I_{\rm LPP} =0$. This action defines a CG in six dimensions which admits Einstein spacetimes in its solutions space, as it has been shown in Section~\ref{ConformalRenormalization}.

\section{Equivalence between the two LPP CG actions}
\label{AppB}

We initially consider the 6D CG action given by Lu, Pang and Pope that admits Schwarzschild-AdS black hole as a solution, that is~\cite{Lu:2013hx},
\begin{equation}
\Lag_{\rm LPP} =\alpha \left (4I_{\rm CG}^{(1)} +I_{\rm CG}^{(2)} -\frac{1}{3}I_{\rm CG}^{(3)}\right ) = \frac{4\alpha }{3}\left (2I_{\rm CG}^{(1)} +I_{\rm CG}^{(2)}\right ) +\frac{\alpha }{3}\left (4I_{\rm CG}^{(1)} -I_{\rm CG}^{(2)}\right ) -\frac{\alpha }{3}I_{\rm CG}^{(3)} \,.
\end{equation}
Taking into account Eq.~\eqref{Pfaffian}, that allows us to write the Pfaffian of the Weyl tensor as a linear combination of the $I_{\rm CG}^{(1)}$ and $I_{\rm CG}^{(2)}$ conformal invariants, we get
\begin{equation}
\Lag_{\rm LPP}
=\frac{\alpha }{4 !}\delta _{\nu _{1}\ldots \nu _{6}}^{\mu _{1}\ldots \mu _{6}}W_{\mu _{1}\mu _{2}}^{\nu _{1}\nu _{2}}W_{\mu _{3}\mu _{4}}^{\nu _{3}\nu _{4}}W_{\mu _{5}\mu _{6}}^{\nu _{5}\nu _{6}} +\frac{\alpha }{3}\left (4I_{\rm CG}^{(1)} -I_{\rm CG}^{(2)}\right ) -\frac{\alpha }{3}I_{\rm CG}^{(3)} \,. \label{LCGpang}
\end{equation}
In the second term, one recognizes a certain combination between $I_{\rm CG}^{(1)}$ and $I_{\rm CG}^{(2)}$, which arises from the Bianchi identity of the Weyl tensor~\eqref{Bianchi}, previously derived in Ref.~\cite{Osborn:2015rna}. Introducing this identity in Eq.~\eqref{LCGpang} one gets
\begin{gather}
\Lag_{\rm LPP} =\alpha \left[\frac{1}{4 !}\delta _{\nu _{1}\ldots \nu _{6}}^{\mu _{1}\ldots \mu _{6}}W_{\mu _{1}\mu _{2}}^{\nu _{1}\nu _{2}}W_{\mu _{3}\mu _{4}}^{\nu _{3}\nu _{4}}W_{\mu _{5}\mu _{6}}^{\nu _{5}\nu _{6}} +\frac{1}{3}W_{\nu \mu }^{\kappa \lambda } \nabla ^{2}W_{\kappa \lambda }^{\nu \mu } -\frac{2}{3}R_{\kappa }^{\omega }W_{\nu \mu }^{\kappa \lambda }W_{\omega \lambda }^{\nu \mu } +\frac{1}{15}RW_{\nu \mu }^{\kappa \lambda }W_{\kappa \lambda }^{\nu \mu } \nonumber \right.\\
\left. -\frac{1}{15}RW_{\nu \mu }^{\kappa \lambda }W_{\kappa \lambda }^{\nu \mu } +8C^{\mu \nu \lambda }C_{\mu \nu \lambda } +\frac{8}{3} \nabla _{\omega }\left (W^{\omega \mu \nu \lambda }C_{\mu \nu \lambda }\right ) -\frac{1}{3}W_{\nu \mu }^{\kappa \lambda } \nabla ^{2}W_{\kappa \lambda }^{\nu \mu } -\frac{4}{3}R_{\kappa }^{\omega }W_{\nu \mu }^{\kappa \lambda }W_{\omega \lambda }^{\nu \mu } \nonumber  \right.\\
\left.+\frac{2}{5}RW_{\nu \mu }^{\kappa \lambda }W_{\kappa \lambda }^{\nu \mu } -\frac{1}{3} \nabla _{\mu }V^{\mu }\right] ,
\end{gather}
where the Schouten tensor has been defined in Eq.~\eqref{BachCottonSchouten}; here we are considering the $D=6$ case. After some simplifications, we obtain
\begin{gather}
\Lag_{\rm LPP } =\alpha \left[\frac{1}{4 !}\delta _{\nu _{1}\ldots \nu _{6}}^{\mu _{1}\ldots \mu _{6}}W_{\mu _{1}\mu _{2}}^{\nu _{1}\nu _{2}}W_{\mu _{3}\mu _{4}}^{\nu _{3}\nu _{4}}W_{\mu _{5}\mu _{6}}^{\nu _{5}\nu _{6}} -8S_{\nu }^{\mu }W_{\mu \kappa }^{\sigma \lambda }W_{\sigma \lambda }^{\nu \kappa } +2SW_{\nu \mu }^{\kappa \lambda }W_{\kappa \lambda }^{\nu \mu } \nonumber  \right.\\
\left.+8C^{\mu \nu \lambda }C_{\mu \nu \lambda } +\frac{1}{3} \nabla _{\mu }\left (8W^{\mu \kappa \lambda \nu }C_{\kappa \lambda \nu } -V^{\mu }\right )\right] \,.
\end{gather}
The terms containing the Schouten tensor can be expressed in a more compact form using
\begin{equation}
\delta _{\nu _{1}\ldots \nu _{5}}^{\mu _{1}\ldots \mu _{5}}W_{\mu _{1}\mu _{2}}^{\nu _{1}\nu _{2}}W_{\mu _{3}\mu _{4}}^{\nu _{3}\nu_{4}}S_{\mu _{5}}^{\nu _{5}} = -16S_{\nu }^{\mu }W^{\nu \kappa \lambda \omega }W_{\mu \kappa \lambda \omega } +4SW_{\nu \mu }^{\kappa \lambda }W_{\kappa \lambda }^{\nu \mu } \,.
\end{equation}
Then, one may rewrite $\Lag_{\rm LPP}$ as
\begin{gather}
\Lag_{\rm LPP} =\alpha \left[\frac{1}{4 !}\delta _{\nu _{1}\ldots \nu _{6}}^{\mu _{1}\ldots \mu _{6}}W_{\mu _{1}\mu _{2}}^{\nu _{1}\nu _{2}}W_{\mu _{3}\mu _{4}}^{\nu _{3}\nu _{4}}W_{\mu _{5}\mu _{6}}^{\nu _{5}\nu _{6}} +\frac{1}{2}\delta _{\nu _{1}\ldots \nu _{5}}^{\mu _{1}\ldots \mu _{5}}W_{\mu _{1}\mu _{2}}^{\nu _{1}\nu _{2}}W_{\mu _{3}\mu _{4}}^{\nu _{3}\nu _{4}}S_{\mu _{5}}^{\nu _{5}} \nonumber  \right.\\
\left.+8C^{\mu \nu \lambda }C_{\mu \nu \lambda } +\frac{1}{3} \nabla _{\mu }\left (8W^{\mu \kappa \lambda \nu }C_{\kappa \lambda \nu } -V^{\mu }\right )\right] \,.
\end{gather}
The last term, being a total derivative, contributes with a boundary term in the 6D CG action which now reads
\begin{gather}I_{\rm LPP} =\alpha \int _{M}\diff{}^{6}x\sqrt{|g|}\left [\frac{1}{4 !}\delta _{\nu _{1}\ldots \nu _{6}}^{\mu _{1}\ldots \mu _{6}}W_{\mu _{1}\mu _{2}}^{\nu _{1}\nu _{2}}W_{\mu _{3}\mu _{4}}^{\nu _{3}\nu _{4}}W_{\mu _{5}\mu _{6}}^{\nu _{5}\nu _{6}} +\frac{1}{2}\delta _{\nu _{1}\ldots \nu _{5}}^{\mu _{1}\ldots \mu _{5}}W_{\mu _{1}\mu _{2}}^{\nu _{1}\nu _{2}}W_{\mu _{3}\mu _{4}}^{\nu _{3}\nu _{4}}S_{\mu _{5}}^{\nu _{5}} +8C^{\mu \nu \lambda }C_{\mu \nu \lambda }\right ] \nonumber  \\
 +\frac{\alpha }{3}\int _{ \partial M}\diff{}^{5}x\sqrt{|h|}\,n_{\mu }\Big[8W^{\mu \kappa \lambda \nu }C_{\kappa \lambda \nu } -4R_{\kappa \sigma }^{\mu \lambda } \nabla ^{\nu }R_{\nu \lambda }^{\kappa \sigma } -3R_{\kappa \sigma }^{\nu \lambda } \nabla ^{\mu }R_{\nu \lambda }^{\kappa \sigma } +5R_{\nu }^{\lambda } \nabla ^{\mu }R_{\lambda }^{\nu } \nonumber  \\
 -\frac{1}{2}R \nabla ^{\mu }R +R_{\nu }^{\mu } \nabla ^{\nu }R -2R_{\nu }^{\lambda } \nabla ^{\nu }R_{\lambda }^{\mu }\Big].\end{gather}
After some algebraic manipulation and using the contracted Bianchi identity $\nabla^\mu G_{\mu\nu}=0$ where $G_{\mu\nu} = R_{\mu\nu} - \tfrac{1}{2}g_{\mu\nu}R$ is the Einstein tensor, alongside the definition of the Cotton tensor in Eq.~\eqref{BachCottonSchouten}, the boundary term can be written as
\begin{gather}I_{\rm LPP}^{\rm (bdy)} 
 =\alpha \int _{ \partial M}\diff{}^{5}x\sqrt{|h|}\,n^{\mu }\left (8W_{\mu }^{\ \kappa \lambda \nu }C_{\kappa \lambda \nu } -W_{\nu \sigma }^{\kappa \lambda } \nabla _{\mu }W_{\kappa \lambda }^{\nu \sigma }\right )\,.
 \end{gather}
 Thus, the CG action given by Lu, Pang and Pope obtains a very compact form:
\begin{gather}I_{\rm LPP} =\alpha \int _{M}\diff{}^{6}x\sqrt{|g|}\left [\frac{1}{4 !}\delta _{\nu _{1}\ldots \nu _{6}}^{\mu _{1}\ldots \mu _{6}}W_{\mu _{1}\mu _{2}}^{\nu _{1}\nu _{2}}W_{\mu _{3}\mu _{4}}^{\nu _{3}\nu _{4}}W_{\mu _{5}\mu _{6}}^{\nu _{5}\nu _{6}} +\frac{1}{2}\delta _{\nu _{1}\ldots \nu _{5}}^{\mu _{1}\ldots \mu _{5}}W_{\mu _{1}\mu _{2}}^{\nu _{1}\nu _{2}}W_{\mu _{3}\mu _{4}}^{\nu _{3}\nu _{4}}S_{\mu _{5}}^{\nu _{5}} +8C^{\mu \nu \lambda }C_{\mu \nu \lambda }\right ] \nonumber  \\
 +\alpha \int _{ \partial M}\diff{}^{5}x\sqrt{|h|}\,n^{\mu }\left (8W_{\mu }^{\ \kappa \lambda \nu }C_{\kappa \lambda \nu } -W_{\nu \sigma }^{\kappa \lambda } \nabla _{\mu }W_{\kappa \lambda }^{\nu \sigma }\right ) . \label{6DCGaction}\end{gather}
 This is the six-dimensional conformal gravity action of Ref.~\cite{Lu:2013hx} written in the form of Eq.~\eqref{ICG2}. 

\bibliography{References}

\begin{thebibliography}{54}%
\makeatletter
\providecommand \@ifxundefined [1]{%
 \@ifx{#1\undefined}
}%
\providecommand \@ifnum [1]{%
 \ifnum #1\expandafter \@firstoftwo
 \else \expandafter \@secondoftwo
 \fi
}%
\providecommand \@ifx [1]{%
 \ifx #1\expandafter \@firstoftwo
 \else \expandafter \@secondoftwo
 \fi
}%
\providecommand \natexlab [1]{#1}%
\providecommand \enquote  [1]{``#1''}%
\providecommand \bibnamefont  [1]{#1}%
\providecommand \bibfnamefont [1]{#1}%
\providecommand \citenamefont [1]{#1}%
\providecommand \href@noop [0]{\@secondoftwo}%
\providecommand \href [0]{\begingroup \@sanitize@url \@href}%
\providecommand \@href[1]{\@@startlink{#1}\@@href}%
\providecommand \@@href[1]{\endgroup#1\@@endlink}%
\providecommand \@sanitize@url [0]{\catcode `\\12\catcode `\$12\catcode `\&12\catcode `\#12\catcode `\^12\catcode `\_12\catcode `\%12\relax}%
\providecommand \@@startlink[1]{}%
\providecommand \@@endlink[0]{}%
\providecommand \url  [0]{\begingroup\@sanitize@url \@url }%
\providecommand \@url [1]{\endgroup\@href {#1}{\urlprefix }}%
\providecommand \urlprefix  [0]{URL }%
\providecommand \Eprint [0]{\href }%
\providecommand \doibase [0]{http://dx.doi.org/}%
\providecommand \selectlanguage [0]{\@gobble}%
\providecommand \bibinfo  [0]{\@secondoftwo}%
\providecommand \bibfield  [0]{\@secondoftwo}%
\providecommand \translation [1]{[#1]}%
\providecommand \BibitemOpen [0]{}%
\providecommand \bibitemStop [0]{}%
\providecommand \bibitemNoStop [0]{.\EOS\space}%
\providecommand \EOS [0]{\spacefactor3000\relax}%
\providecommand \BibitemShut  [1]{\csname bibitem#1\endcsname}%
\let\auto@bib@innerbib\@empty
\bibitem [{\citenamefont {Weyl}(1918)}]{Weyl:1918ib}%
  \BibitemOpen
  \bibfield  {author} {\bibinfo {author} {\bibfnamefont {H.}~\bibnamefont {Weyl}},\ }\href@noop {} {\bibfield  {journal} {\bibinfo  {journal} {Sitzungsber. Preuss. Akad. Wiss. Berlin (Math. Phys. )}\ }\textbf {\bibinfo {volume} {1918}},\ \bibinfo {pages} {465} (\bibinfo {year} {1918})}\BibitemShut {NoStop}%
\bibitem [{\citenamefont {Weyl}(1919)}]{Weyl:1919fi}%
  \BibitemOpen
  \bibfield  {author} {\bibinfo {author} {\bibfnamefont {H.}~\bibnamefont {Weyl}},\ }\href {\doibase 10.1002/andp.19193641002} {\bibfield  {journal} {\bibinfo  {journal} {Annalen Phys.}\ }\textbf {\bibinfo {volume} {59}},\ \bibinfo {pages} {101} (\bibinfo {year} {1919})}\BibitemShut {NoStop}%
\bibitem [{\citenamefont {Bach}(1921)}]{Bach:1921}%
  \BibitemOpen
  \bibfield  {author} {\bibinfo {author} {\bibfnamefont {R.}~\bibnamefont {Bach}},\ }\href {\doibase 10.1007/BF01378338} {\bibfield  {journal} {\bibinfo  {journal} {Mathematische Zeitschrift}\ }\textbf {\bibinfo {volume} {9}},\ \bibinfo {pages} {110} (\bibinfo {year} {1921})}\BibitemShut {NoStop}%
\bibitem [{\citenamefont {Stelle}(1977)}]{Stelle:1976gc}%
  \BibitemOpen
  \bibfield  {author} {\bibinfo {author} {\bibfnamefont {K.~S.}\ \bibnamefont {Stelle}},\ }\href {\doibase 10.1103/PhysRevD.16.953} {\bibfield  {journal} {\bibinfo  {journal} {Phys. Rev. D}\ }\textbf {\bibinfo {volume} {16}},\ \bibinfo {pages} {953} (\bibinfo {year} {1977})}\BibitemShut {NoStop}%
\bibitem [{\citenamefont {Capper}\ and\ \citenamefont {Duff}(1975)}]{Capper:1975ig}%
  \BibitemOpen
  \bibfield  {author} {\bibinfo {author} {\bibfnamefont {D.~M.}\ \bibnamefont {Capper}}\ and\ \bibinfo {author} {\bibfnamefont {M.~J.}\ \bibnamefont {Duff}},\ }\href {\doibase 10.1016/0375-9601(75)90030-4} {\bibfield  {journal} {\bibinfo  {journal} {Phys. Lett. A}\ }\textbf {\bibinfo {volume} {53}},\ \bibinfo {pages} {361} (\bibinfo {year} {1975})}\BibitemShut {NoStop}%
\bibitem [{\citenamefont {Fradkin}\ and\ \citenamefont {Tseytlin}(1982)}]{Fradkin:1981iu}%
  \BibitemOpen
  \bibfield  {author} {\bibinfo {author} {\bibfnamefont {E.~S.}\ \bibnamefont {Fradkin}}\ and\ \bibinfo {author} {\bibfnamefont {A.~A.}\ \bibnamefont {Tseytlin}},\ }\href {\doibase 10.1016/0550-3213(82)90444-8} {\bibfield  {journal} {\bibinfo  {journal} {Nucl. Phys. B}\ }\textbf {\bibinfo {volume} {201}},\ \bibinfo {pages} {469} (\bibinfo {year} {1982})}\BibitemShut {NoStop}%
\bibitem [{\citenamefont {Julve}\ and\ \citenamefont {Tonin}(1978)}]{Julve:1978xn}%
  \BibitemOpen
  \bibfield  {author} {\bibinfo {author} {\bibfnamefont {J.}~\bibnamefont {Julve}}\ and\ \bibinfo {author} {\bibfnamefont {M.}~\bibnamefont {Tonin}},\ }\href {\doibase 10.1007/BF02748637} {\bibfield  {journal} {\bibinfo  {journal} {Nuovo Cim. B}\ }\textbf {\bibinfo {volume} {46}},\ \bibinfo {pages} {137} (\bibinfo {year} {1978})}\BibitemShut {NoStop}%
\bibitem [{\citenamefont {Mannheim}\ and\ \citenamefont {Kazanas}(1989)}]{Mannheim:1988dj}%
  \BibitemOpen
  \bibfield  {author} {\bibinfo {author} {\bibfnamefont {P.~D.}\ \bibnamefont {Mannheim}}\ and\ \bibinfo {author} {\bibfnamefont {D.}~\bibnamefont {Kazanas}},\ }\href {\doibase 10.1086/167623} {\bibfield  {journal} {\bibinfo  {journal} {Astrophys. J.}\ }\textbf {\bibinfo {volume} {342}},\ \bibinfo {pages} {635} (\bibinfo {year} {1989})}\BibitemShut {NoStop}%
\bibitem [{\citenamefont {Mannheim}(2006)}]{Mannheim:2005bfa}%
  \BibitemOpen
  \bibfield  {author} {\bibinfo {author} {\bibfnamefont {P.~D.}\ \bibnamefont {Mannheim}},\ }\href {\doibase 10.1016/j.ppnp.2005.08.001} {\bibfield  {journal} {\bibinfo  {journal} {Prog. Part. Nucl. Phys.}\ }\textbf {\bibinfo {volume} {56}},\ \bibinfo {pages} {340} (\bibinfo {year} {2006})},\ \Eprint {http://arxiv.org/abs/astro-ph/0505266} {arXiv:astro-ph/0505266} \BibitemShut {NoStop}%
\bibitem [{\citenamefont {Mannheim}\ and\ \citenamefont {O'Brien}(2011)}]{Mannheim:2010ti}%
  \BibitemOpen
  \bibfield  {author} {\bibinfo {author} {\bibfnamefont {P.~D.}\ \bibnamefont {Mannheim}}\ and\ \bibinfo {author} {\bibfnamefont {J.~G.}\ \bibnamefont {O'Brien}},\ }\href {\doibase 10.1103/PhysRevLett.106.121101} {\bibfield  {journal} {\bibinfo  {journal} {Phys. Rev. Lett.}\ }\textbf {\bibinfo {volume} {106}},\ \bibinfo {pages} {121101} (\bibinfo {year} {2011})},\ \Eprint {http://arxiv.org/abs/1007.0970} {arXiv:1007.0970 [astro-ph.CO]} \BibitemShut {NoStop}%
\bibitem [{\citenamefont {Mannheim}(2012)}]{Mannheim:2011ds}%
  \BibitemOpen
  \bibfield  {author} {\bibinfo {author} {\bibfnamefont {P.~D.}\ \bibnamefont {Mannheim}},\ }\href {\doibase 10.1007/s10701-011-9608-6} {\bibfield  {journal} {\bibinfo  {journal} {Found. Phys.}\ }\textbf {\bibinfo {volume} {42}},\ \bibinfo {pages} {388} (\bibinfo {year} {2012})},\ \Eprint {http://arxiv.org/abs/1101.2186} {arXiv:1101.2186 [hep-th]} \BibitemShut {NoStop}%
\bibitem [{\citenamefont {Berkovits}\ and\ \citenamefont {Witten}(2004)}]{Berkovits:2004jj}%
  \BibitemOpen
  \bibfield  {author} {\bibinfo {author} {\bibfnamefont {N.}~\bibnamefont {Berkovits}}\ and\ \bibinfo {author} {\bibfnamefont {E.}~\bibnamefont {Witten}},\ }\href {\doibase 10.1088/1126-6708/2004/08/009} {\bibfield  {journal} {\bibinfo  {journal} {JHEP}\ }\textbf {\bibinfo {volume} {08}},\ \bibinfo {pages} {009} (\bibinfo {year} {2004})}\BibitemShut {NoStop}%
\bibitem [{\citenamefont {Kaku}\ and\ \citenamefont {Townsend}(1978)}]{Kaku:1978ea}%
  \BibitemOpen
  \bibfield  {author} {\bibinfo {author} {\bibfnamefont {M.}~\bibnamefont {Kaku}}\ and\ \bibinfo {author} {\bibfnamefont {P.}~\bibnamefont {Townsend}},\ }\href {\doibase 10.1016/0370-2693(78)90098-9} {\bibfield  {journal} {\bibinfo  {journal} {Phys. Lett. B}\ }\textbf {\bibinfo {volume} {76}},\ \bibinfo {pages} {54} (\bibinfo {year} {1978})}\BibitemShut {NoStop}%
\bibitem [{\citenamefont {Kaku}\ \emph {et~al.}(1978)\citenamefont {Kaku}, \citenamefont {Townsend},\ and\ \citenamefont {van Nieuwenhuizen}}]{Kaku:1978nz}%
  \BibitemOpen
  \bibfield  {author} {\bibinfo {author} {\bibfnamefont {M.}~\bibnamefont {Kaku}}, \bibinfo {author} {\bibfnamefont {P.}~\bibnamefont {Townsend}}, \ and\ \bibinfo {author} {\bibfnamefont {P.}~\bibnamefont {van Nieuwenhuizen}},\ }\href {\doibase 10.1103/PhysRevD.17.3179} {\bibfield  {journal} {\bibinfo  {journal} {Phys. Rev. D}\ }\textbf {\bibinfo {volume} {17}},\ \bibinfo {pages} {3179} (\bibinfo {year} {1978})}\BibitemShut {NoStop}%
\bibitem [{\citenamefont {Kaku}\ \emph {et~al.}(1977)\citenamefont {Kaku}, \citenamefont {Townsend},\ and\ \citenamefont {van Nieuwenhuizen}}]{Kaku:1977pa}%
  \BibitemOpen
  \bibfield  {author} {\bibinfo {author} {\bibfnamefont {M.}~\bibnamefont {Kaku}}, \bibinfo {author} {\bibfnamefont {P.~K.}\ \bibnamefont {Townsend}}, \ and\ \bibinfo {author} {\bibfnamefont {P.}~\bibnamefont {van Nieuwenhuizen}},\ }\href {\doibase 10.1016/0370-2693(77)90552-4} {\bibfield  {journal} {\bibinfo  {journal} {Phys. Lett. B}\ }\textbf {\bibinfo {volume} {69}},\ \bibinfo {pages} {304} (\bibinfo {year} {1977})}\BibitemShut {NoStop}%
\bibitem [{\citenamefont {Fradkin}\ and\ \citenamefont {Tseytlin}(1985)}]{Fradkin:1985am}%
  \BibitemOpen
  \bibfield  {author} {\bibinfo {author} {\bibfnamefont {E.~S.}\ \bibnamefont {Fradkin}}\ and\ \bibinfo {author} {\bibfnamefont {A.~A.}\ \bibnamefont {Tseytlin}},\ }\href {\doibase 10.1016/0370-1573(85)90138-3} {\bibfield  {journal} {\bibinfo  {journal} {Phys. Rept.}\ }\textbf {\bibinfo {volume} {119}},\ \bibinfo {pages} {233} (\bibinfo {year} {1985})}\BibitemShut {NoStop}%
\bibitem [{\citenamefont {de~Wit}\ \emph {et~al.}(1981)\citenamefont {de~Wit}, \citenamefont {van Holten},\ and\ \citenamefont {Van~Proeyen}}]{deWit:1980lyi}%
  \BibitemOpen
  \bibfield  {author} {\bibinfo {author} {\bibfnamefont {B.}~\bibnamefont {de~Wit}}, \bibinfo {author} {\bibfnamefont {J.~W.}\ \bibnamefont {van Holten}}, \ and\ \bibinfo {author} {\bibfnamefont {A.}~\bibnamefont {Van~Proeyen}},\ }\href {\doibase 10.1016/0550-3213(83)90548-5} {\bibfield  {journal} {\bibinfo  {journal} {Nucl. Phys. B}\ }\textbf {\bibinfo {volume} {184}},\ \bibinfo {pages} {77} (\bibinfo {year} {1981})},\ \bibinfo {note} {[Erratum: Nucl.Phys.B 222, 516 (1983)]}\BibitemShut {NoStop}%
\bibitem [{\citenamefont {Bergshoeff}\ \emph {et~al.}(1981)\citenamefont {Bergshoeff}, \citenamefont {de~Roo},\ and\ \citenamefont {de~Wit}}]{Bergshoeff:1980is}%
  \BibitemOpen
  \bibfield  {author} {\bibinfo {author} {\bibfnamefont {E.}~\bibnamefont {Bergshoeff}}, \bibinfo {author} {\bibfnamefont {M.}~\bibnamefont {de~Roo}}, \ and\ \bibinfo {author} {\bibfnamefont {B.}~\bibnamefont {de~Wit}},\ }\href {\doibase 10.1016/0550-3213(81)90465-X} {\bibfield  {journal} {\bibinfo  {journal} {Nucl. Phys. B}\ }\textbf {\bibinfo {volume} {182}},\ \bibinfo {pages} {173} (\bibinfo {year} {1981})}\BibitemShut {NoStop}%
\bibitem [{\citenamefont {Liu}\ and\ \citenamefont {Tseytlin}(1998)}]{Liu:1998bu}%
  \BibitemOpen
  \bibfield  {author} {\bibinfo {author} {\bibfnamefont {H.}~\bibnamefont {Liu}}\ and\ \bibinfo {author} {\bibfnamefont {A.~A.}\ \bibnamefont {Tseytlin}},\ }\href {\doibase 10.1016/S0550-3213(98)00443-X} {\bibfield  {journal} {\bibinfo  {journal} {Nucl. Phys. B}\ }\textbf {\bibinfo {volume} {533}},\ \bibinfo {pages} {88} (\bibinfo {year} {1998})},\ \Eprint {http://arxiv.org/abs/hep-th/9804083} {arXiv:hep-th/9804083} \BibitemShut {NoStop}%
\bibitem [{\citenamefont {Ferrara}\ \emph {et~al.}(2018)\citenamefont {Ferrara}, \citenamefont {Kehagias},\ and\ \citenamefont {L\"ust}}]{Ferrara:2018wqd}%
  \BibitemOpen
  \bibfield  {author} {\bibinfo {author} {\bibfnamefont {S.}~\bibnamefont {Ferrara}}, \bibinfo {author} {\bibfnamefont {A.}~\bibnamefont {Kehagias}}, \ and\ \bibinfo {author} {\bibfnamefont {D.}~\bibnamefont {L\"ust}},\ }\href {\doibase 10.1007/JHEP08(2018)197} {\bibfield  {journal} {\bibinfo  {journal} {JHEP}\ }\textbf {\bibinfo {volume} {08}},\ \bibinfo {pages} {197} (\bibinfo {year} {2018})},\ \Eprint {http://arxiv.org/abs/1806.10016} {arXiv:1806.10016 [hep-th]} \BibitemShut {NoStop}%
\bibitem [{\citenamefont {Andrianopoli}\ and\ \citenamefont {D'Auria}(2014)}]{Andrianopoli:2014aqa}%
  \BibitemOpen
  \bibfield  {author} {\bibinfo {author} {\bibfnamefont {L.}~\bibnamefont {Andrianopoli}}\ and\ \bibinfo {author} {\bibfnamefont {R.}~\bibnamefont {D'Auria}},\ }\href {\doibase 10.1007/JHEP08(2014)012} {\bibfield  {journal} {\bibinfo  {journal} {JHEP}\ }\textbf {\bibinfo {volume} {08}},\ \bibinfo {pages} {012} (\bibinfo {year} {2014})},\ \Eprint {http://arxiv.org/abs/1405.2010} {arXiv:1405.2010 [hep-th]} \BibitemShut {NoStop}%
\bibitem [{\citenamefont {D'Auria}\ and\ \citenamefont {Ravera}(2021)}]{DAuria:2021dth}%
  \BibitemOpen
  \bibfield  {author} {\bibinfo {author} {\bibfnamefont {R.}~\bibnamefont {D'Auria}}\ and\ \bibinfo {author} {\bibfnamefont {L.}~\bibnamefont {Ravera}},\ }\href {\doibase 10.1103/PhysRevD.104.084034} {\bibfield  {journal} {\bibinfo  {journal} {Phys. Rev. D}\ }\textbf {\bibinfo {volume} {104}},\ \bibinfo {pages} {084034} (\bibinfo {year} {2021})},\ \Eprint {http://arxiv.org/abs/2101.10978} {arXiv:2101.10978 [hep-th]} \BibitemShut {NoStop}%
\bibitem [{\citenamefont {Ferrara}\ \emph {et~al.}(2020)\citenamefont {Ferrara}, \citenamefont {Kehagias},\ and\ \citenamefont {L\"ust}}]{Ferrara:2020zef}%
  \BibitemOpen
  \bibfield  {author} {\bibinfo {author} {\bibfnamefont {S.}~\bibnamefont {Ferrara}}, \bibinfo {author} {\bibfnamefont {A.}~\bibnamefont {Kehagias}}, \ and\ \bibinfo {author} {\bibfnamefont {D.}~\bibnamefont {L\"ust}},\ }in\ \href@noop {} {\emph {\bibinfo {booktitle} {{57th International School of Subnuclear Physics}: {In Search for the Unexpected}}}}\ (\bibinfo {year} {2020})\ \Eprint {http://arxiv.org/abs/2001.04998} {arXiv:2001.04998 [hep-th]} \BibitemShut {NoStop}%
\bibitem [{\citenamefont {Chamseddine}\ and\ \citenamefont {Connes}(1997)}]{Chamseddine:1996zu}%
  \BibitemOpen
  \bibfield  {author} {\bibinfo {author} {\bibfnamefont {A.~H.}\ \bibnamefont {Chamseddine}}\ and\ \bibinfo {author} {\bibfnamefont {A.}~\bibnamefont {Connes}},\ }\href {\doibase 10.1007/s002200050126} {\bibfield  {journal} {\bibinfo  {journal} {Commun. Math. Phys.}\ }\textbf {\bibinfo {volume} {186}},\ \bibinfo {pages} {731} (\bibinfo {year} {1997})},\ \Eprint {http://arxiv.org/abs/hep-th/9606001} {arXiv:hep-th/9606001} \BibitemShut {NoStop}%
\bibitem [{\citenamefont {Manolakos}\ \emph {et~al.}(2020)\citenamefont {Manolakos}, \citenamefont {Manousselis},\ and\ \citenamefont {Zoupanos}}]{Manolakos:2019fle}%
  \BibitemOpen
  \bibfield  {author} {\bibinfo {author} {\bibfnamefont {G.}~\bibnamefont {Manolakos}}, \bibinfo {author} {\bibfnamefont {P.}~\bibnamefont {Manousselis}}, \ and\ \bibinfo {author} {\bibfnamefont {G.}~\bibnamefont {Zoupanos}},\ }\href {\doibase 10.1007/JHEP08(2020)001} {\bibfield  {journal} {\bibinfo  {journal} {JHEP}\ }\textbf {\bibinfo {volume} {08}},\ \bibinfo {pages} {001} (\bibinfo {year} {2020})},\ \Eprint {http://arxiv.org/abs/1902.10922} {arXiv:1902.10922 [hep-th]} \BibitemShut {NoStop}%
\bibitem [{\citenamefont {Manolakos}\ \emph {et~al.}(2021)\citenamefont {Manolakos}, \citenamefont {Manousselis},\ and\ \citenamefont {Zoupanos}}]{Manolakos:2021rcl}%
  \BibitemOpen
  \bibfield  {author} {\bibinfo {author} {\bibfnamefont {G.}~\bibnamefont {Manolakos}}, \bibinfo {author} {\bibfnamefont {P.}~\bibnamefont {Manousselis}}, \ and\ \bibinfo {author} {\bibfnamefont {G.}~\bibnamefont {Zoupanos}},\ }\href {\doibase 10.1002/prop.202100085} {\bibfield  {journal} {\bibinfo  {journal} {Fortsch. Phys.}\ }\textbf {\bibinfo {volume} {69}},\ \bibinfo {pages} {2100085} (\bibinfo {year} {2021})},\ \Eprint {http://arxiv.org/abs/2104.13746} {arXiv:2104.13746 [hep-th]} \BibitemShut {NoStop}%
\bibitem [{\citenamefont {Ostrogradsky}(1850)}]{Ostrogradsky:1850fid}%
  \BibitemOpen
  \bibfield  {author} {\bibinfo {author} {\bibfnamefont {M.}~\bibnamefont {Ostrogradsky}},\ }\href@noop {} {\bibfield  {journal} {\bibinfo  {journal} {Mem. Acad. St. Petersbourg}\ }\textbf {\bibinfo {volume} {6}},\ \bibinfo {pages} {385} (\bibinfo {year} {1850})}\BibitemShut {NoStop}%
\bibitem [{\citenamefont {Riegert}(1984)}]{Riegert:1984hf}%
  \BibitemOpen
  \bibfield  {author} {\bibinfo {author} {\bibfnamefont {R.~J.}\ \bibnamefont {Riegert}},\ }\href {\doibase 10.1016/0375-9601(84)90648-0} {\bibfield  {journal} {\bibinfo  {journal} {Phys. Lett. A}\ }\textbf {\bibinfo {volume} {105}},\ \bibinfo {pages} {110} (\bibinfo {year} {1984})}\BibitemShut {NoStop}%
\bibitem [{\citenamefont {Mannheim}(2022)}]{Mannheim:2021oat}%
  \BibitemOpen
  \bibfield  {author} {\bibinfo {author} {\bibfnamefont {P.~D.}\ \bibnamefont {Mannheim}},\ }\href {\doibase 10.1393/ncc/i2022-22027-6} {\bibfield  {journal} {\bibinfo  {journal} {Nuovo Cim. C}\ }\textbf {\bibinfo {volume} {45}},\ \bibinfo {pages} {27} (\bibinfo {year} {2022})},\ \Eprint {http://arxiv.org/abs/2109.12743} {arXiv:2109.12743 [hep-th]} \BibitemShut {NoStop}%
\bibitem [{\citenamefont {Maldacena}(2011)}]{Maldacena:2011mk}%
  \BibitemOpen
  \bibfield  {author} {\bibinfo {author} {\bibfnamefont {J.}~\bibnamefont {Maldacena}},\ }\href@noop {} {\  (\bibinfo {year} {2011})},\ \Eprint {http://arxiv.org/abs/1105.5632} {arXiv:1105.5632 [hep-th]} \BibitemShut {NoStop}%
\bibitem [{\citenamefont {Grumiller}\ \emph {et~al.}(2014)\citenamefont {Grumiller}, \citenamefont {Irakleidou}, \citenamefont {Lovrekovic},\ and\ \citenamefont {McNees}}]{Grumiller:2013mxa}%
  \BibitemOpen
  \bibfield  {author} {\bibinfo {author} {\bibfnamefont {D.}~\bibnamefont {Grumiller}}, \bibinfo {author} {\bibfnamefont {M.}~\bibnamefont {Irakleidou}}, \bibinfo {author} {\bibfnamefont {I.}~\bibnamefont {Lovrekovic}}, \ and\ \bibinfo {author} {\bibfnamefont {R.}~\bibnamefont {McNees}},\ }\href {\doibase 10.1103/PhysRevLett.112.111102} {\bibfield  {journal} {\bibinfo  {journal} {Phys. Rev. Lett.}\ }\textbf {\bibinfo {volume} {112}},\ \bibinfo {pages} {111102} (\bibinfo {year} {2014})}\BibitemShut {NoStop}%
\bibitem [{\citenamefont {Fefferman}\ and\ \citenamefont {Graham}(1985)}]{AST_1985__S131__95_0}%
  \BibitemOpen
  \bibfield  {author} {\bibinfo {author} {\bibfnamefont {C.}~\bibnamefont {Fefferman}}\ and\ \bibinfo {author} {\bibfnamefont {C.~R.}\ \bibnamefont {Graham}}\ }(\bibinfo  {publisher} {Soci\'et\'e math\'ematique de France},\ \bibinfo {year} {1985})\ pp.\ \bibinfo {pages} {95--116}\BibitemShut {NoStop}%
\bibitem [{\citenamefont {Hell}\ \emph {et~al.}(2023)\citenamefont {Hell}, \citenamefont {Lust},\ and\ \citenamefont {Zoupanos}}]{Hell:2023rbf}%
  \BibitemOpen
  \bibfield  {author} {\bibinfo {author} {\bibfnamefont {A.}~\bibnamefont {Hell}}, \bibinfo {author} {\bibfnamefont {D.}~\bibnamefont {Lust}}, \ and\ \bibinfo {author} {\bibfnamefont {G.}~\bibnamefont {Zoupanos}},\ }\href@noop {} {\  (\bibinfo {year} {2023})},\ \Eprint {http://arxiv.org/abs/2306.13714} {arXiv:2306.13714 [hep-th]} \BibitemShut {NoStop}%
\bibitem [{\citenamefont {Anastasiou}\ and\ \citenamefont {Olea}(2016)}]{Anastasiou:2016jix}%
  \BibitemOpen
  \bibfield  {author} {\bibinfo {author} {\bibfnamefont {G.}~\bibnamefont {Anastasiou}}\ and\ \bibinfo {author} {\bibfnamefont {R.}~\bibnamefont {Olea}},\ }\href {\doibase 10.1103/PhysRevD.94.086008} {\bibfield  {journal} {\bibinfo  {journal} {Phys. Rev. D}\ }\textbf {\bibinfo {volume} {94}},\ \bibinfo {pages} {086008} (\bibinfo {year} {2016})}\BibitemShut {NoStop}%
\bibitem [{\citenamefont {MacDowell}\ and\ \citenamefont {Mansouri}(1977)}]{MacDowell:1977jt}%
  \BibitemOpen
  \bibfield  {author} {\bibinfo {author} {\bibfnamefont {S.~W.}\ \bibnamefont {MacDowell}}\ and\ \bibinfo {author} {\bibfnamefont {F.}~\bibnamefont {Mansouri}},\ }\href {\doibase 10.1103/PhysRevLett.38.739} {\bibfield  {journal} {\bibinfo  {journal} {Phys. Rev. Lett.}\ }\textbf {\bibinfo {volume} {38}},\ \bibinfo {pages} {739} (\bibinfo {year} {1977})},\ \bibinfo {note} {[Erratum: Phys. Rev. Lett. 38, 1376 (1977)]}\BibitemShut {NoStop}%
\bibitem [{\citenamefont {Miskovic}\ and\ \citenamefont {Olea}(2009)}]{Miskovic:2009bm}%
  \BibitemOpen
  \bibfield  {author} {\bibinfo {author} {\bibfnamefont {O.}~\bibnamefont {Miskovic}}\ and\ \bibinfo {author} {\bibfnamefont {R.}~\bibnamefont {Olea}},\ }\href {\doibase 10.1103/PhysRevD.79.124020} {\bibfield  {journal} {\bibinfo  {journal} {Phys. Rev. D}\ }\textbf {\bibinfo {volume} {79}},\ \bibinfo {pages} {124020} (\bibinfo {year} {2009})}\BibitemShut {NoStop}%
\bibitem [{\citenamefont {Anastasiou}\ \emph {et~al.}(2021{\natexlab{a}})\citenamefont {Anastasiou}, \citenamefont {Araya},\ and\ \citenamefont {Olea}}]{Anastasiou:2020mik}%
  \BibitemOpen
  \bibfield  {author} {\bibinfo {author} {\bibfnamefont {G.}~\bibnamefont {Anastasiou}}, \bibinfo {author} {\bibfnamefont {I.~J.}\ \bibnamefont {Araya}}, \ and\ \bibinfo {author} {\bibfnamefont {R.}~\bibnamefont {Olea}},\ }\href {\doibase 10.1007/JHEP01(2021)134} {\bibfield  {journal} {\bibinfo  {journal} {JHEP}\ }\textbf {\bibinfo {volume} {01}},\ \bibinfo {pages} {134} (\bibinfo {year} {2021}{\natexlab{a}})}\BibitemShut {NoStop}%
\bibitem [{\citenamefont {L\"u}\ \emph {et~al.}(2013)\citenamefont {L\"u}, \citenamefont {Pang},\ and\ \citenamefont {Pope}}]{Lu:2013hx}%
  \BibitemOpen
  \bibfield  {author} {\bibinfo {author} {\bibfnamefont {H.}~\bibnamefont {L\"u}}, \bibinfo {author} {\bibfnamefont {Y.}~\bibnamefont {Pang}}, \ and\ \bibinfo {author} {\bibfnamefont {C.~N.}\ \bibnamefont {Pope}},\ }\href {\doibase 10.1103/PhysRevD.87.104013} {\bibfield  {journal} {\bibinfo  {journal} {Phys. Rev. D}\ }\textbf {\bibinfo {volume} {87}},\ \bibinfo {pages} {104013} (\bibinfo {year} {2013})},\ \Eprint {http://arxiv.org/abs/1301.7083} {arXiv:1301.7083 [hep-th]} \BibitemShut {NoStop}%
\bibitem [{\citenamefont {Olea}(2005)}]{Olea:2005gb}%
  \BibitemOpen
  \bibfield  {author} {\bibinfo {author} {\bibfnamefont {R.}~\bibnamefont {Olea}},\ }\href {\doibase 10.1088/1126-6708/2005/06/023} {\bibfield  {journal} {\bibinfo  {journal} {JHEP}\ }\textbf {\bibinfo {volume} {06}},\ \bibinfo {pages} {023} (\bibinfo {year} {2005})}\BibitemShut {NoStop}%
\bibitem [{\citenamefont {de~Haro}\ \emph {et~al.}(2001)\citenamefont {de~Haro}, \citenamefont {Solodukhin},\ and\ \citenamefont {Skenderis}}]{deHaro:2000vlm}%
  \BibitemOpen
  \bibfield  {author} {\bibinfo {author} {\bibfnamefont {S.}~\bibnamefont {de~Haro}}, \bibinfo {author} {\bibfnamefont {S.~N.}\ \bibnamefont {Solodukhin}}, \ and\ \bibinfo {author} {\bibfnamefont {K.}~\bibnamefont {Skenderis}},\ }\href {\doibase 10.1007/s002200100381} {\bibfield  {journal} {\bibinfo  {journal} {Commun. Math. Phys.}\ }\textbf {\bibinfo {volume} {217}},\ \bibinfo {pages} {595} (\bibinfo {year} {2001})},\ \Eprint {http://arxiv.org/abs/hep-th/0002230} {arXiv:hep-th/0002230} \BibitemShut {NoStop}%
\bibitem [{\citenamefont {Henningson}\ and\ \citenamefont {Skenderis}(1998)}]{Henningson:1998gx}%
  \BibitemOpen
  \bibfield  {author} {\bibinfo {author} {\bibfnamefont {M.}~\bibnamefont {Henningson}}\ and\ \bibinfo {author} {\bibfnamefont {K.}~\bibnamefont {Skenderis}},\ }\href {\doibase 10.1088/1126-6708/1998/07/023} {\bibfield  {journal} {\bibinfo  {journal} {JHEP}\ }\textbf {\bibinfo {volume} {07}},\ \bibinfo {pages} {023} (\bibinfo {year} {1998})},\ \Eprint {http://arxiv.org/abs/hep-th/9806087} {arXiv:hep-th/9806087} \BibitemShut {NoStop}%
\bibitem [{\citenamefont {Anastasiou}\ \emph {et~al.}(2023)\citenamefont {Anastasiou}, \citenamefont {Araya}, \citenamefont {Busnego-Barrientos}, \citenamefont {Corral},\ and\ \citenamefont {Merino}}]{Anastasiou:2022wjq}%
  \BibitemOpen
  \bibfield  {author} {\bibinfo {author} {\bibfnamefont {G.}~\bibnamefont {Anastasiou}}, \bibinfo {author} {\bibfnamefont {I.~J.}\ \bibnamefont {Araya}}, \bibinfo {author} {\bibfnamefont {M.}~\bibnamefont {Busnego-Barrientos}}, \bibinfo {author} {\bibfnamefont {C.}~\bibnamefont {Corral}}, \ and\ \bibinfo {author} {\bibfnamefont {N.}~\bibnamefont {Merino}},\ }\href {\doibase 10.1103/PhysRevD.107.104049} {\bibfield  {journal} {\bibinfo  {journal} {Phys. Rev. D}\ }\textbf {\bibinfo {volume} {107}},\ \bibinfo {pages} {104049} (\bibinfo {year} {2023})},\ \Eprint {http://arxiv.org/abs/2212.04364} {arXiv:2212.04364 [hep-th]} \BibitemShut {NoStop}%
\bibitem [{\citenamefont {Anastasiou}\ \emph {et~al.}(2018)\citenamefont {Anastasiou}, \citenamefont {Araya}, \citenamefont {Arias},\ and\ \citenamefont {Olea}}]{Anastasiou:2018mfk}%
  \BibitemOpen
  \bibfield  {author} {\bibinfo {author} {\bibfnamefont {G.}~\bibnamefont {Anastasiou}}, \bibinfo {author} {\bibfnamefont {I.~J.}\ \bibnamefont {Araya}}, \bibinfo {author} {\bibfnamefont {C.}~\bibnamefont {Arias}}, \ and\ \bibinfo {author} {\bibfnamefont {R.}~\bibnamefont {Olea}},\ }\href {\doibase 10.1007/JHEP08(2018)136} {\bibfield  {journal} {\bibinfo  {journal} {JHEP}\ }\textbf {\bibinfo {volume} {08}},\ \bibinfo {pages} {136} (\bibinfo {year} {2018})},\ \Eprint {http://arxiv.org/abs/1806.10708} {arXiv:1806.10708 [hep-th]} \BibitemShut {NoStop}%
\bibitem [{\citenamefont {Anastasiou}\ \emph {et~al.}(2019)\citenamefont {Anastasiou}, \citenamefont {Araya}, \citenamefont {Guijosa},\ and\ \citenamefont {Olea}}]{Anastasiou:2019ldc}%
  \BibitemOpen
  \bibfield  {author} {\bibinfo {author} {\bibfnamefont {G.}~\bibnamefont {Anastasiou}}, \bibinfo {author} {\bibfnamefont {I.~J.}\ \bibnamefont {Araya}}, \bibinfo {author} {\bibfnamefont {A.}~\bibnamefont {Guijosa}}, \ and\ \bibinfo {author} {\bibfnamefont {R.}~\bibnamefont {Olea}},\ }\href {\doibase 10.1007/JHEP10(2019)221} {\bibfield  {journal} {\bibinfo  {journal} {JHEP}\ }\textbf {\bibinfo {volume} {10}},\ \bibinfo {pages} {221} (\bibinfo {year} {2019})},\ \Eprint {http://arxiv.org/abs/1908.11447} {arXiv:1908.11447 [hep-th]} \BibitemShut {NoStop}%
\bibitem [{\citenamefont {Anastasiou}\ \emph {et~al.}(2020)\citenamefont {Anastasiou}, \citenamefont {Miskovic}, \citenamefont {Olea},\ and\ \citenamefont {Papadimitriou}}]{Anastasiou:2020zwc}%
  \BibitemOpen
  \bibfield  {author} {\bibinfo {author} {\bibfnamefont {G.}~\bibnamefont {Anastasiou}}, \bibinfo {author} {\bibfnamefont {O.}~\bibnamefont {Miskovic}}, \bibinfo {author} {\bibfnamefont {R.}~\bibnamefont {Olea}}, \ and\ \bibinfo {author} {\bibfnamefont {I.}~\bibnamefont {Papadimitriou}},\ }\href {\doibase 10.1007/JHEP08(2020)061} {\bibfield  {journal} {\bibinfo  {journal} {JHEP}\ }\textbf {\bibinfo {volume} {08}},\ \bibinfo {pages} {061} (\bibinfo {year} {2020})}\BibitemShut {NoStop}%
\bibitem [{\citenamefont {Araya}\ \emph {et~al.}(2021)\citenamefont {Araya}, \citenamefont {Edelstein}, \citenamefont {Sanchez}, \citenamefont {Rodriguez},\ and\ \citenamefont {Lopez}}]{Araya:2021atx}%
  \BibitemOpen
  \bibfield  {author} {\bibinfo {author} {\bibfnamefont {I.~J.}\ \bibnamefont {Araya}}, \bibinfo {author} {\bibfnamefont {J.~D.}\ \bibnamefont {Edelstein}}, \bibinfo {author} {\bibfnamefont {A.~R.}\ \bibnamefont {Sanchez}}, \bibinfo {author} {\bibfnamefont {D.~V.}\ \bibnamefont {Rodriguez}}, \ and\ \bibinfo {author} {\bibfnamefont {A.~V.}\ \bibnamefont {Lopez}},\ }\href {\doibase 10.1007/JHEP09(2021)142} {\bibfield  {journal} {\bibinfo  {journal} {JHEP}\ }\textbf {\bibinfo {volume} {09}},\ \bibinfo {pages} {142} (\bibinfo {year} {2021})},\ \Eprint {http://arxiv.org/abs/2108.01126} {arXiv:2108.01126 [hep-th]} \BibitemShut {NoStop}%
\bibitem [{\citenamefont {Eguchi}\ \emph {et~al.}(1980)\citenamefont {Eguchi}, \citenamefont {Gilkey},\ and\ \citenamefont {Hanson}}]{Eguchi:1980jx}%
  \BibitemOpen
  \bibfield  {author} {\bibinfo {author} {\bibfnamefont {T.}~\bibnamefont {Eguchi}}, \bibinfo {author} {\bibfnamefont {P.~B.}\ \bibnamefont {Gilkey}}, \ and\ \bibinfo {author} {\bibfnamefont {A.~J.}\ \bibnamefont {Hanson}},\ }\href {\doibase 10.1016/0370-1573(80)90130-1} {\bibfield  {journal} {\bibinfo  {journal} {Phys. Rept.}\ }\textbf {\bibinfo {volume} {66}},\ \bibinfo {pages} {213} (\bibinfo {year} {1980})}\BibitemShut {NoStop}%
\bibitem [{\citenamefont {Skenderis}(2002)}]{Skenderis:2002wp}%
  \BibitemOpen
  \bibfield  {author} {\bibinfo {author} {\bibfnamefont {K.}~\bibnamefont {Skenderis}},\ }\href {\doibase 10.1088/0264-9381/19/22/306} {\bibfield  {journal} {\bibinfo  {journal} {Class. Quant. Grav.}\ }\textbf {\bibinfo {volume} {19}},\ \bibinfo {pages} {5849} (\bibinfo {year} {2002})},\ \Eprint {http://arxiv.org/abs/hep-th/0209067} {arXiv:hep-th/0209067} \BibitemShut {NoStop}%
\bibitem [{\citenamefont {Papadimitriou}\ and\ \citenamefont {Skenderis}(2005)}]{Papadimitriou:2004ap}%
  \BibitemOpen
  \bibfield  {author} {\bibinfo {author} {\bibfnamefont {I.}~\bibnamefont {Papadimitriou}}\ and\ \bibinfo {author} {\bibfnamefont {K.}~\bibnamefont {Skenderis}},\ }\href {\doibase 10.4171/013-1/4} {\bibfield  {journal} {\bibinfo  {journal} {IRMA Lect. Math. Theor. Phys.}\ }\textbf {\bibinfo {volume} {8}},\ \bibinfo {pages} {73} (\bibinfo {year} {2005})},\ \Eprint {http://arxiv.org/abs/hep-th/0404176} {arXiv:hep-th/0404176} \BibitemShut {NoStop}%
\bibitem [{\citenamefont {Balasubramanian}\ and\ \citenamefont {Kraus}(1999)}]{Balasubramanian:1999re}%
  \BibitemOpen
  \bibfield  {author} {\bibinfo {author} {\bibfnamefont {V.}~\bibnamefont {Balasubramanian}}\ and\ \bibinfo {author} {\bibfnamefont {P.}~\bibnamefont {Kraus}},\ }\href {\doibase 10.1007/s002200050764} {\bibfield  {journal} {\bibinfo  {journal} {Commun. Math. Phys.}\ }\textbf {\bibinfo {volume} {208}},\ \bibinfo {pages} {413} (\bibinfo {year} {1999})}\BibitemShut {NoStop}%
\bibitem [{\citenamefont {Emparan}\ \emph {et~al.}(1999)\citenamefont {Emparan}, \citenamefont {Johnson},\ and\ \citenamefont {Myers}}]{Emparan:1999pm}%
  \BibitemOpen
  \bibfield  {author} {\bibinfo {author} {\bibfnamefont {R.}~\bibnamefont {Emparan}}, \bibinfo {author} {\bibfnamefont {C.~V.}\ \bibnamefont {Johnson}}, \ and\ \bibinfo {author} {\bibfnamefont {R.~C.}\ \bibnamefont {Myers}},\ }\href {\doibase 10.1103/PhysRevD.60.104001} {\bibfield  {journal} {\bibinfo  {journal} {Phys. Rev. D}\ }\textbf {\bibinfo {volume} {60}},\ \bibinfo {pages} {104001} (\bibinfo {year} {1999})}\BibitemShut {NoStop}%
\bibitem [{\citenamefont {Iyer}\ and\ \citenamefont {Wald}(1994)}]{Iyer:1994ys}%
  \BibitemOpen
  \bibfield  {author} {\bibinfo {author} {\bibfnamefont {V.}~\bibnamefont {Iyer}}\ and\ \bibinfo {author} {\bibfnamefont {R.~M.}\ \bibnamefont {Wald}},\ }\href {\doibase 10.1103/PhysRevD.50.846} {\bibfield  {journal} {\bibinfo  {journal} {Phys. Rev. D}\ }\textbf {\bibinfo {volume} {50}},\ \bibinfo {pages} {846} (\bibinfo {year} {1994})}\BibitemShut {NoStop}%
\bibitem [{\citenamefont {Anastasiou}\ \emph {et~al.}(2021{\natexlab{b}})\citenamefont {Anastasiou}, \citenamefont {Araya}, \citenamefont {Corral},\ and\ \citenamefont {Olea}}]{Anastasiou:2021tlv}%
  \BibitemOpen
  \bibfield  {author} {\bibinfo {author} {\bibfnamefont {G.}~\bibnamefont {Anastasiou}}, \bibinfo {author} {\bibfnamefont {I.~J.}\ \bibnamefont {Araya}}, \bibinfo {author} {\bibfnamefont {C.}~\bibnamefont {Corral}}, \ and\ \bibinfo {author} {\bibfnamefont {R.}~\bibnamefont {Olea}},\ }\href {\doibase 10.1007/JHEP07(2021)156} {\bibfield  {journal} {\bibinfo  {journal} {JHEP}\ }\textbf {\bibinfo {volume} {2021}},\ \bibinfo {pages} {156} (\bibinfo {year} {2021}{\natexlab{b}})}\BibitemShut {NoStop}%
\bibitem [{\citenamefont {Osborn}\ and\ \citenamefont {Stergiou}(2015)}]{Osborn:2015rna}%
  \BibitemOpen
  \bibfield  {author} {\bibinfo {author} {\bibfnamefont {H.}~\bibnamefont {Osborn}}\ and\ \bibinfo {author} {\bibfnamefont {A.}~\bibnamefont {Stergiou}},\ }\href {\doibase 10.1007/JHEP04(2015)157} {\bibfield  {journal} {\bibinfo  {journal} {JHEP}\ }\textbf {\bibinfo {volume} {04}},\ \bibinfo {pages} {157} (\bibinfo {year} {2015})},\ \Eprint {http://arxiv.org/abs/1501.01308} {arXiv:1501.01308 [hep-th]} \BibitemShut {NoStop}%
\end{thebibliography}%

\end{document}